\documentclass[preprint, superscriptaddress, preprintnumbers, amsmath, amssymb]{revtex4}

\allowdisplaybreaks
\allowdisplaybreaks[4]
\usepackage{CJK}
\usepackage{graphicx}
\usepackage{dcolumn}
\usepackage{bm}
\usepackage[dvipdfm,
            pdfstartview=FitH,
            CJKbookmarks=true,
            bookmarksnumbered=true,
            bookmarksopen=true,
            colorlinks=true,
            pdfborder=001,
            linkcolor=blue,
            anchorcolor=blue,
            citecolor=blue
            ]{hyperref}

\begin{document}

\begin{CJK}{GBK}{song}

\title{Behavior of the collective rotor in nuclear chiral motion}

\author{Q. B. Chen}\email{qbchen@pku.edu.cn}
\affiliation{Physik-Department, Technische Universit\"{a}t
M\"{u}nchen, D-85747 Garching, Germany}

\author{N. Kaiser}\email{nkaiser@ph.tum.de}
\affiliation{Physik-Department, Technische Universit\"{a}t
M\"{u}nchen, D-85747 Garching, Germany}

\author{Ulf-G. Mei{\ss}ner}\email{meissner@hiskp.uni-bonn.de}
\affiliation{Helmholtz-Institut f\"{u}r Strahlen- und Kernphysik and
Bethe Center for Theoretical Physics, Universit\"{a}t Bonn, D-53115
Bonn, Germany}

\affiliation{Institute for Advanced Simulation, Institut f\"{u}r
Kernphysik and J\"{u}lich Center for Hadron Physics,
Forschungszentrum J\"{u}lich, D-52425 J\"{u}lich, Germany}

\affiliation{Ivane Javakhishvili Tbilisi State University, 0186
Tbilisi, Georgia}

\author{J. Meng}\email{mengj@pku.edu.cn}
\affiliation{State Key Laboratory of Nuclear Physics and Technology,
             School of Physics, Peking University, Beijing 100871, China}%
\affiliation{Yukawa Institute for Theoretical Physics, Kyoto
             University, Kyoto 606-8502, Japan}

\date{\today}

\begin{abstract}

The behavior of the collective rotor in the chiral motion of triaxially deformed
nuclei is investigated using the particle rotor model by transforming the wave
functions from the $K$-representation to the $R$-representation.
After examining the energy spectra of the doublet bands and their
energy differences as functions of the triaxial deformation,
the angular momentum components of the rotor, proton, neutron, and the total system
are investigated. Moreover, the probability distributions of the rotor
angular momentum ($R$-plots) and their projections onto the three principal
axes ($K_R$-plots) are analyzed. The evolution of the chiral mode from
a chiral vibration at the low spins to a chiral rotation at high spins
is illustrated at triaxial deformations $\gamma=20^\circ$ and $30^\circ$.

\end{abstract}

\maketitle


\section{Introduction}
\label{intro}

Nuclear chiral rotation is an exotic form of spontaneous symmetry
breaking. It can occur when high-$j$ proton states (particles) lie above
the Fermi level and high-$j$ neutron states (holes) lie below the Fermi
level (or vice versa), and at the same time the nuclear core is of triaxial ellipsoidal
shape~\cite{Frauendorf1997NPA}. The angular momenta of the valence particles
and holes are aligned along the short and long axes of the triaxial core, respectively,
while the angular momentum of the rotational core is aligned along the intermediate
axis. The three angular momenta can be arranged to form a left-handed or
a right-handed system. Such an arrangement leads to the breaking of chiral
symmetry ($\chi=\mathcal{T}\mathcal{R}_2(\pi)$, with
time reversal $\mathcal{T}$ and $180^\circ$ degree rotation $\mathcal{R}_2(\pi)$)
in the body-fixed frame. With the restoration of this symmetry in the
laboratory frame, degenerate doublet $\Delta I = 1$ bands
with the same parity, so-called chiral doublet bands~\cite{Frauendorf1997NPA},
occur.

So far, more than 50 candidates for this phenomenon have been observed in odd-odd
nuclei as well as in odd-$A$ and even-even nuclei, and these are spread over the
mass regions $A \sim 80$, 100, 130, and 190. For more details, see
the review articles~\cite{J.Meng2010JPG, J.Meng2014IJMPE, Bark2014IJMPE, J.Meng2016PS,
Raduta2016PPNP, Frauendorf2018PS} and the corresponding data
tables in Ref.~\cite{B.W.Xiong2019ADNDT}. With the prediction~\cite{J.Meng2006PRC}
and confirmation~\cite{Ayangeakaa2013PRL} of multiple chiral
doublets (M$\chi$D) in a single nucleus, the investigation of
nuclear chirality continues to be one of the hottest topic in modern
nuclear physics~\cite{J.Peng2008PRC, J.M.Yao2009PRC, J.Li2011PRC,
Droste2009EPJA, Q.B.Chen2010PRC, Hamamoto2013PRC, Tonev2014PRL,
Lieder2014PRL, Rather2014PRL, Kuti2014PRL, C.Liu2016PRL,
H.Zhang2016CPC, Grodner2018PRL, J.Li2018PRC, Petrache2018PRC,
Q.B.Chen2018PRC, Moon2018PLB, Q.B.Chen2018PLB, Roy2018PLB, B.Qi2018PRC,
J.Peng2018PRC, Q.B.Chen2018PRC_v1, Bujor2018PRC}.

By now it is well-known that chiral rotations (aplanar rotations of the total
angular momentum) can exist only above a critical spin $I$, see Refs.~\cite{Olbratowski2004PRL,
Olbratowski2006PRC, P.W.Zhao2017PLB, Grodner2018PRL, Q.B.Chen2018PRC_v1}. Actually,
at low spins a chiral vibration, understood as an oscillation of the total angular
momentum between the left- and the right-handed configuration, happens.
This suggests that the orientation of the angular momenta of the rotor, the particle(s),
and the hole(s) are coplanar near the bandhead of a chiral band. This feature is
caused by the fact that the angular momentum of the rotor is much smaller than those of
the proton and the neutron near the bandhead~\cite{Q.B.Chen2018PRC_v1}. On the other
hand, at high spin, a chiral rotation occurs, which is driven by
the increase of the rotor angular momentum along the intermediate axis.

Obviously, the rotor plays an essential role in the evolution of the chiral mode
from the chiral vibration to the chiral rotation. Therefore, the detailed exploration
of the behavior of the collective rotor in nuclear chiral motion is of high interest.
Previously, such investigations were mainly carried out by calculating expectation
values of components of the rotor angular momentum~\cite{S.Q.Zhang2007PRC, B.Qi2009PLB,
B.Qi2009PRC, Lawrie2010PLB, Q.B.Chen2010PRC, Hamamoto2013PRC, H.Zhang2016CPC,
Petrache2016PRC}). Only rare attempts have been made to investigate the
detailed wave functions of the collective rotor in chiral bands. To our knowledge
only in Ref.~\cite{Droste2009EPJA}, the rotor wave functions have been explored at
the beginning and the end of chiral bands.

In this work we will take the system of one $h_{11/2}$ proton-particle and one
$h_{11/2}$ neutron-hole coupled to a triaxial rigid rotor as concrete example to
investigate systematically the behavior of the collective rotor angular momentum
in nuclear chiral motion.

Among various nuclear models, the particle rotor model (PRM) has
been widely used to describe chiral doublet bands with different
kinds of particle-hole configurations~\cite{Frauendorf1997NPA, J.Peng2003PRC, Koike2004PRL,
B.Qi2009PRC, H.Zhang2016CPC, Q.B.Chen2018PRC, Q.B.Chen2018PRC_v1, Koike2003PRC,
S.Q.Zhang2007PRC, S.Y.Wang2007PRC, S.Y.Wang2008PRC, Lawrie2008PRC,
Lawrie2010PLB, Shirinda2012EPJA, B.Qi2009PLB, B.Qi2011PRC, Ayangeakaa2013PRL,
B.Qi2013PRC, Lieder2014PRL, Kuti2014PRL, Petrache2016PRC, Q.B.Chen2018PLB}.
It is a quantum mechanical model, which treats the collective rotation and the intrinsic
single-particle motions based on a description of the system in the laboratory
frame. The pertinent Hamiltonian is diagonalized with total angular momentum $I$
as a good quantum number, and the energy splittings and tunneling probabilities
between doublet bands can be obtained directly from eigenvalues and eigenfunctions.

Usually, the PRM Hamiltonian is diagonalized in the strong coupling basis~\cite{Bohr1975,
Ring1980book}, where the projection of the total spin onto the 3-axis of the
intrinsic frame is a good quantum number, denoted by $K$. In this $K$-representation,
the rotor angular momentum $R$ and its three possible projections $K_R$ on
the intrinsic axes do not appear explicitly. In order to give the proper
wave function of the rotor, one has to express the PRM wave function in terms
of the weak-coupling basis~\cite{Bohr1975, Ring1980book}, in which both $R$
and $K_R$ are good quantum numbers. This transformation gives the $R$-representation,
and from the corresponding probability distributions one can derive
the {\it $R$-plot} and three {\it $K_R$-plots}.

The technique to transform the PRM wave function from the $K$-representation
to the $R$-representation is outlined in the textbook~\cite{Bohr1975}.
In particular, we have used it in Ref.~\cite{Streck2018PRC} to investigate
the behavior of the collective rotor in the wobbling motion of $^{135}$Pr.
In the present work, we extend the same method to investigate chiral
bands based on a two-quasiparticle configuration.


\section{Theoretical framework}

\subsection{Particle rotor Hamiltonian}

In the particle rotor model (PRM) the Hamiltonian for a system with one
proton and one neutron coupled to a triaxial rigid rotor is composed
as~\cite{Frauendorf1997NPA, J.Peng2003PRC, Koike2004PRL, S.Q.Zhang2007PRC, B.Qi2009PLB}
\begin{align}\label{eq1}
\hat{H}_\textrm{PRM}=\hat{H}_{\rm coll}+\hat{H}_{p}+\hat{H}_{n},
\end{align}
where $\hat{H}_{\rm coll}$ represents the Hamiltonian of the rigid rotor,
\begin{align}\label{eq6}
\hat{H}_{\rm coll}
 &=\sum_{k=1}^3 \frac{\hat{R}_k^2}{2\mathcal{J}_k}\\
\label{eq11}
 &=\sum_{k=1}^3 \frac{(\hat{I}_k-\hat{j}_{pk}-\hat{j}_{nk})^2}{2\mathcal{J}_k},
\end{align}
with the index $k=1$, 2, 3 denoting the three principal axes in the
body-fixed frame. Here, $\hat{R}_k$ and $\hat{I}_k$ are the angular
momentum operators of the collective rotor and the total nucleus,
while $\hat{j}_{p(n)k}$ is the angular momentum operator of the valence
proton (neutron). Moreover, the parameters $\mathcal{J}_k$ are the three
principal moments of inertia. When calculating matrix elements of
$\hat{H}_{\rm coll}$, the $R$-representation is most conveniently used because
its form in Eq.~(\ref{eq6}), while the form in Eq.~(\ref{eq11}) is preferably
treated in the $K$-representation.

The Hamiltonians $\hat{H}_p$ and $\hat{H}_n$ describe the single
proton and neutron outside of the rotor. For a nucleon in a single-$j$ shell
orbital, it is given by
\begin{align}\label{eq2}
\hat{H}_{p(n)}=\pm \frac{1}{2}C\Big\{\cos \gamma\big[\hat{j}_3^2-\frac{j(j+1)}{3}\big]
              +\frac{\sin \gamma}{2\sqrt{3}}\big(\hat{j}_+^2+\hat{j}_-^2\big)\Big\},
\end{align}
where the plus sign refers to a particle and the minus sign to a hole
and $\gamma$ serves as the triaxial deformation parameter.
The coupling parameter $C$ is proportional to the quadrupole
deformation parameter $\beta$ of the rotor.

\subsection{Basis transformation from $K$-representation
to $R$-representation}

The PRM Hamiltonian in Eq.~(\ref{eq1}) is usually solved by diagonalization in
the strong-coupling basis ($K$-representation)~\cite{Bohr1975, Ring1980book}
\begin{align}\label{eq3}
 &\quad |j_p\Omega_p j_n\Omega_n K,IM\rangle \notag\\
 &=\sqrt{\frac{1}{2}}\Big[
 |j_p\Omega_p\rangle|j_n\Omega_n\rangle|IMK\rangle\notag\\
 &+(-1)^{I-j_p-j_n}|j_p-\Omega_p\rangle|j_n-\Omega_n\rangle|IM-K\rangle\Big].
\end{align}
where $I$ denotes the total angular momentum quantum number of
the odd-odd nuclear system (rotor plus proton and neutron) and $M$ refers to the
projection onto the 3-axis of the laboratory frame. Furthermore, $\Omega_{p(n)}$
is the quantum number for the 3-axis component of the valence nucleon angular
momentum operator $\bm{j}_{p(n)}$ in the intrinsic frame, while $D_{MK}^I(\bm{\omega})$
are the usual Wigner-functions, depending on three Euler angles $\bm{\omega}=(\psi^\prime,
\theta^\prime,\phi^\prime)$. Under the requirement of the $\textrm{D}_2$ symmetry of a triaxial
nucleus~\cite{Bohr1975}, $K$ and $\Omega_{p}$ take the values: $K=-I,\dots,I$
and $\Omega_p=-j_p, \dots, j_p$. The quantum number $\Omega_n$ goes over the range
$\Omega_n=-j_n, \dots, j_n$ and it has to fulfil the condition that $K_R=K-\Omega_p-\Omega_n$ is
a positive even integer. In the special case $K_R=0$, only positive values
$\Omega_n=1/2, \dots, j_n$ are allowed. With these choices, the dimension
of the Hamiltonian matrix is $(2I+1)(2j_p+1)(2j_n+1)/4$.

In the $K$-representation (\ref{eq3}), the rotor angular momentum quantum number
$R$ does not appear explicitly. In order to obtain the wave function of the rotor
in the $R$-representation, one has to change the basis. The details of
this orthogonal transformation for a triaxial system with odd particle number can
be found in Refs.~\cite{Davids2004PRC, Modi2017PRC, Streck2018PRC}. Here, following
the procedure presented in Ref.~\cite{Streck2018PRC}, we will extend it here to
an odd-odd nucleus.

The rotational wave function of the total nuclear system in the laboratory frame
can be expressed in the $R$-representation as
\begin{align}\label{eq4}
 &\quad |j_pj_n(J_{pn})RK_R,IM\rangle \notag\\
 &=\sum_{M_{pn}, M_R} \langle J_{pn}M_{pn}RM_R|IM\rangle
 |J_{pn}M_{pn}\rangle  |R M_R \tau \rangle \notag\\
 &=\sum_{M_{pn},M_R,m_p,m_n} \langle J_{pn}M_{pn}RM_R|IM\rangle \notag\\
 &\times \langle j_p m_p j_n m_n|J_{pn}M_{pn}\rangle
 |j_pm_p\rangle  |j_n m_n\rangle  |R M_R \tau\rangle,
\end{align}
where first the coupling of $\bm{j}_p$ and $\bm{j}_n$ to $\bm{J}_{pn}$ is performed
and after that $\bm{J}_{pn}$ and the rotor quantum number $\bm{R}$ are coupled to total
angular momentum $\bm{I}$. In the above expression, $M_R$, $M_{pn}$, and $m_{p(n)}$ are the
projection quantum numbers of $\bm{R}$, $\bm{J}_{pn}$,
and $\bm{j}_{p(n)}$ on the 3-axis in the laboratory frame, respectively.
Obviously, the appearance of Clebsch-Gordan coefficients requires $M=M_{pn}+M_R=
m_p+m_n+M_R$. The value of $J_{pn}$  is in the range
$|j_p-j_n| \leq J_{pn} \leq j_p+j_n$. Accordingly, for a
given $J_{pn}$, the value of $R$ must satisfy the triangular
condition $|I-J_{pn}| \leq R \leq I+J_{pn}$ of angular momentum coupling.
The additional quantum number $\tau$ refers to the projection of $\bm{R}$
onto a specific body-fixed axis.

Now we perform the transformation from the $R$-representation to the
$K$-representation. In the $K$-representation, the quantum number $\tau$ is
identified with the projection $K_R$ of $\bm{R}$ onto a principal axis.
Making use of Wigner-functions, the wave functions of the two particles
and the rotor in Eq.~(\ref{eq4}) can be written as
\begin{align}
\label{eq5}
 |j_p m_p\rangle &=\sum_{\Omega_p=-j_p}^{j_p}
  D_{m_p\Omega_p}^{j_p}(\bm{\omega})|j_p\Omega_p\rangle,\\
\label{eq7}
 |j_n m_n\rangle &=\sum_{\Omega_n=-j_n}^{j_n}
  D_{m_n\Omega_n}^{j_n}(\bm{\omega})|j_n\Omega_n\rangle,\\
\label{eq8}
 |RM_R K_R\rangle &=\sqrt{\frac{2R+1}{16\pi^2(1+\delta_{K_R0})}}\notag\\
 &\quad \times\Big[D_{M_RK_R}^R(\bm{\omega})+(-1)^R D_{M_R-K_R}^R(\bm{\omega})\Big],
\end{align}
where, as mentioned already above, $K_R$ is an even integer ranging from $0$ to $R$,
with $K_R=0$ is excluded for odd $R$. Both restrictions come from the $\textrm{D}_2$
symmetry of a triaxial nucleus~\cite{Bohr1975}.

Substituting Eqs.~(\ref{eq5})-(\ref{eq8}) into Eq.~(\ref{eq4}), one obtains
\begin{align}
 &\quad |j_pj_n(J_{pn})RK_R,IM\rangle \notag\\
 &=\sum_{K,\Omega_p,\Omega_n}
 A_{j_p\Omega_p j_n\Omega_n,RK_R}^{IK, J_{pn}\Omega_{pn}}
  |j_p\Omega_p j_n\Omega_n K,IM\rangle,
\end{align}
with the expansion coefficients
\begin{align}
 &A_{j_p\Omega_p j_n\Omega_n,RK_R}^{IK, J_{pn}\Omega_{pn}} \notag\\
 &=\sqrt{\frac{2R+1}{2I+1}}\sqrt{1+\delta_{K_R,0}}\notag\\
 &\quad \times \langle j_p\Omega_p j_n\Omega_n|J_{pn}\Omega_{pn}\rangle
 \langle J_{pn}\Omega_{pn} R K_R|IK\rangle.
\end{align}
Obviously, the transformation between the $K$-representation and
the $R$-representation is an orthogonal transformation, and therefore the
expansion coefficients satisfy the relations
\begin{align}
 &\quad \sum_{K, \Omega_p, \Omega_n, \Omega_{pn}} A_{j_p\Omega_p j_n\Omega_n,RK_R}^{IK, J_{pn}\Omega_{pn}}
   A_{j_p\Omega_p j_n\Omega_n,R^\prime K_R^\prime}^{IK, J_{pn}\Omega_{pn}} \notag\\
  &=\delta_{RR^\prime}\delta_{K_R K_{R}^\prime},\\
 &\quad \sum_{R, K_R, J_{pn}, \Omega_{pn}} A_{j_p\Omega_p j_n\Omega_n,RK_R}^{IK, J_{pn}\Omega_{pn}}
   A_{j_p\Omega_p^\prime j_n\Omega_n^\prime, RK_R}^{IK^\prime, J_{pn}\Omega_{pn}} \notag\\
  &=\delta_{\Omega_p\Omega_p^\prime} \delta_{\Omega_n\Omega_n^\prime} \delta_{KK^\prime}.
\end{align}
Due to the orthogonality property, the inverse transformation
follows immediately as
\begin{align}\label{eq9}
 &\quad |j_p\Omega_p j_n\Omega_n K,IM\rangle \notag\\
 &=\sum_{R,K_R,J_{pn},\Omega_{pn}}
  A_{j_p\Omega_p j_n\Omega_n,RK_R}^{IK, J_{pn}\Omega_{pn}}
  |j_pj_n(J_{pn})RK_R,IM\rangle.
\end{align}
With this formula, we have successfully transformed the PRM basis
functions from the $K$-representation to the $R$-representation.

The advantage of the basis states in Eq.~(\ref{eq9}) is a more convenient
calculation of the matrix elements of the collective rotor Hamiltonian
\begin{align}
 &\quad \langle j_p\Omega_p^\prime j_n\Omega_n^\prime K^\prime,IM|
  \hat{H}_{\textrm{coll}}|j_p\Omega_p j_n\Omega_n K,IM\rangle\notag\\
 &=\sum_{R^\prime, K_R^\prime, J_{pn}^\prime, \Omega_{pn}^\prime}
   \sum_{R, K_R, J_{pn}, \Omega_{pn}}
 A_{j_p\Omega_p^\prime j_n\Omega_n^\prime,
  R^\prime K_R^\prime}^{IK^\prime, J_{pn}^\prime \Omega_{pn}^\prime}
 A_{j_p\Omega_p j_n\Omega_n,RK_R}^{IK, J_{pn}\Omega_{pn}}\notag\\
 &\times \langle j_pj_n(J_{pn}^\prime)R K_R^\prime,IM|\hat{H}_{\textrm{coll}}
 |j_pj_n(J_{pn})RK_R,IM\rangle \notag\\
 &=\sum_{R, K_R, K_R^\prime, J_{pn}, \Omega_{pn}, \Omega_{pn}^\prime}
 A_{j_p\Omega_p j_n\Omega_n,RK_R^\prime}^{IK^\prime, J_{pn} \Omega_{pn}^\prime}
 A_{j_p\Omega_p j_n\Omega_n,RK_R}^{IK, J_{pn}\Omega_{pn}}\notag\\
 &\times \Big(\sum_i c_{K_R^\prime}^{Ri} E_{Ri} c_{K_R}^{Ri}\Big).
\end{align}
The energy eigenvalues $E_{Ri}$ and corresponding expansion coefficients
$c_{K_R}^{Ri}$ ($i$ labels the different eigenstates) are obtained by
diagonalizing the collective rotor Hamiltonian $\hat{H}_{\textrm{coll}}$
in the basis $|RM_RK_R\rangle$ introduced in Eq.~(\ref{eq8})
\begin{align}
 \hat{H}_{\textrm{coll}}|RM_Ri\rangle &= E_{Ri}|RM_Ri\rangle,\\
 |RM_Ri\rangle &=\sum_{K_R} c_{K_R}^{Ri} |RM_RK_R\rangle.
\end{align}

Most importantly, the transformation (\ref{eq9}) allows us also to calculate
the probability distributions of the rotor angular momentum, which will
be given in the following subsection.

\subsection{$R$-plots and $K_R$-plots}

With the above preparations, the PRM eigenfunctions can be expressed as
\begin{align}
 &\quad |IM\rangle\notag\\
  &=\sum_{K,\Omega_p,\Omega_n} d_{K,\Omega_p,\Omega_n} |j_p\Omega_p j_n\Omega_n K,IM\rangle\notag\\
  &=\sum_{K,\Omega_p,\Omega_n} d_{K,\Omega_p,\Omega_n} \sum_{R,K_R,J_{pn},\Omega_{pn}}
  A_{j_p\Omega_p j_n\Omega_n,RK_R}^{IK, J_{pn}\Omega_{pn}}\notag\\
  &\times \sum_{M_{pn},M_R,m_p,m_n} \langle J_{pn}M_{pn}RM_R|IM\rangle \notag\\
  &\times \langle j_p m_p j_n m_n|J_{pn}M_{pn}\rangle
  |j_pm_p\rangle  |j_n m_n\rangle |RK_RM_R\rangle,
\end{align}
where the expansion coefficients $d_{K,\Omega_p, \Omega_n}$ are obtained by
diagonalizing the total PRM Hamiltonian in Eq.~(\ref{eq1}). Hence, the
probabilities for given $R$ and $K_R$ are calculated as
\begin{align}
 P_{R, K_R}=\sum_{J_{pn}}\Big(\sum_{K, \Omega_p, \Omega_n, \Omega_{pn}}
   d_{K, \Omega_p, \Omega_n} A_{j_p\Omega_p j_n\Omega_n,RK_R}^{IK, J_{pn}\Omega_{pn}} \Big)^2,
\end{align}
and they satisfy the normalization condition
\begin{align}
 \sum_{R,K_R}P_{R,K_R}=1.
\end{align}
The $R$-plot is obtained from the summed probabilities
\begin{align}
 P_R=\sum_{K_R} P_{R,K_R},
\end{align}
whereas in the $K_R$-plot the probabilities are summed differently
\begin{align}
 P_{K_R}=\sum_R P_{R,K_R}.
\end{align}

Moreover, the expectation value of the squared angular momentum operator
$\hat{R}_3^2$ follows as
\begin{align}
 \langle IM|\hat{R}_3^2|IM\rangle=\sum_{R,K_R} K_R^2 P_{R,K_R}.
\end{align}

\section{Numerical details}

In our calculation, a system of one $h_{11/2}$ proton particle and
one $h_{11/2}$ neutron hole coupled to a triaxial rigid rotor with
quadruple deformation parameters $\beta = 0.23$ and triaxial deformation
parameter $\gamma \in [0^\circ, 30^\circ]$ is considered for the purpose of
illustrating the angular momentum geometry. Moments of inertia of
the irrotational flow type $\mathcal{J}_k=\mathcal{J}_0\sin^2(\gamma-2k\pi/3)$
$(k=1, 2, 3)$ with $\mathcal{J}_0=30~\hbar^2/\textrm{MeV}$ are used.


\section{Results and discussion}

\subsection{Energy spectra}

\begin{figure}[!ht]
  \begin{center}
    \includegraphics[width=6.0 cm]{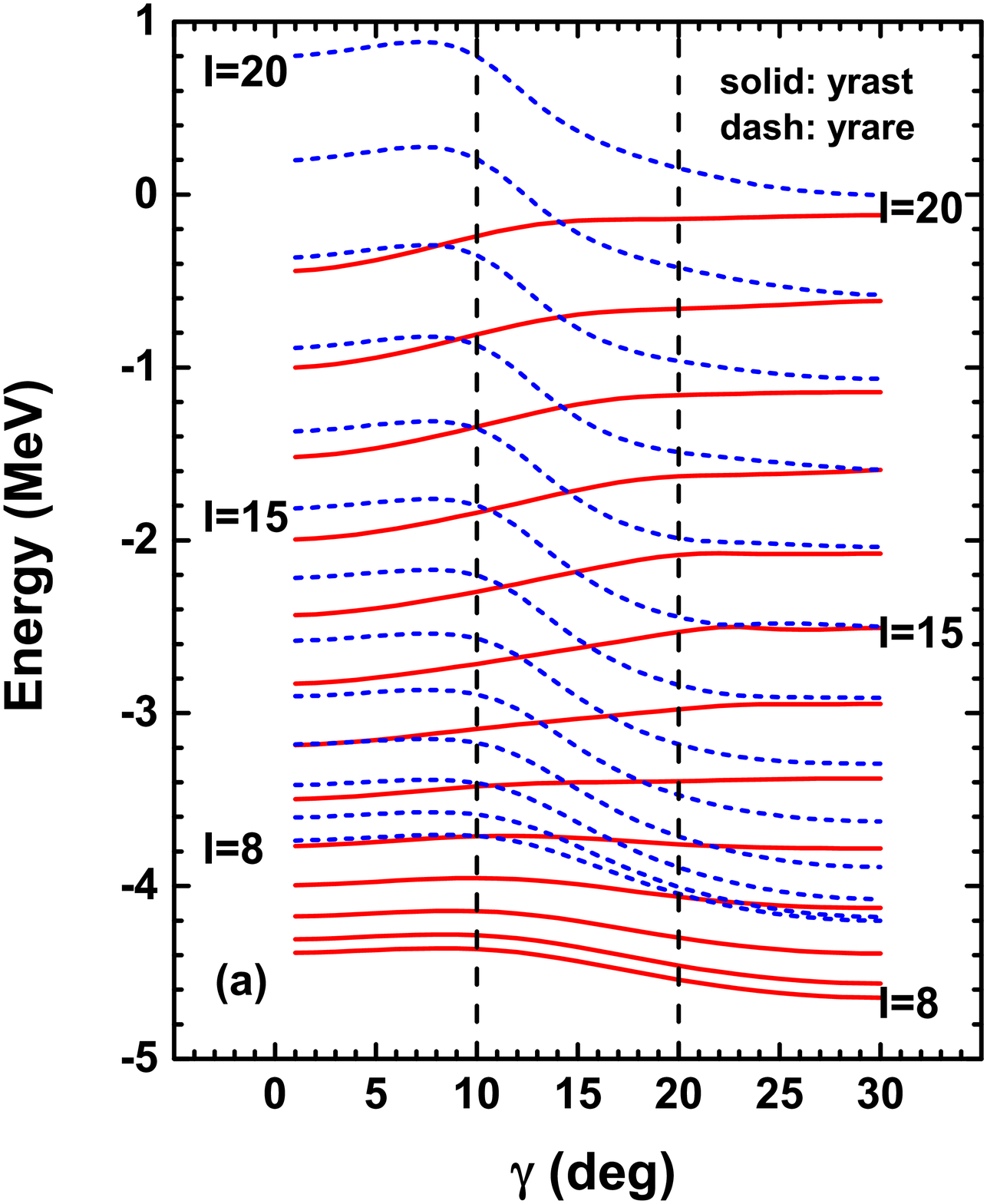}\quad
    \includegraphics[width=6.0 cm]{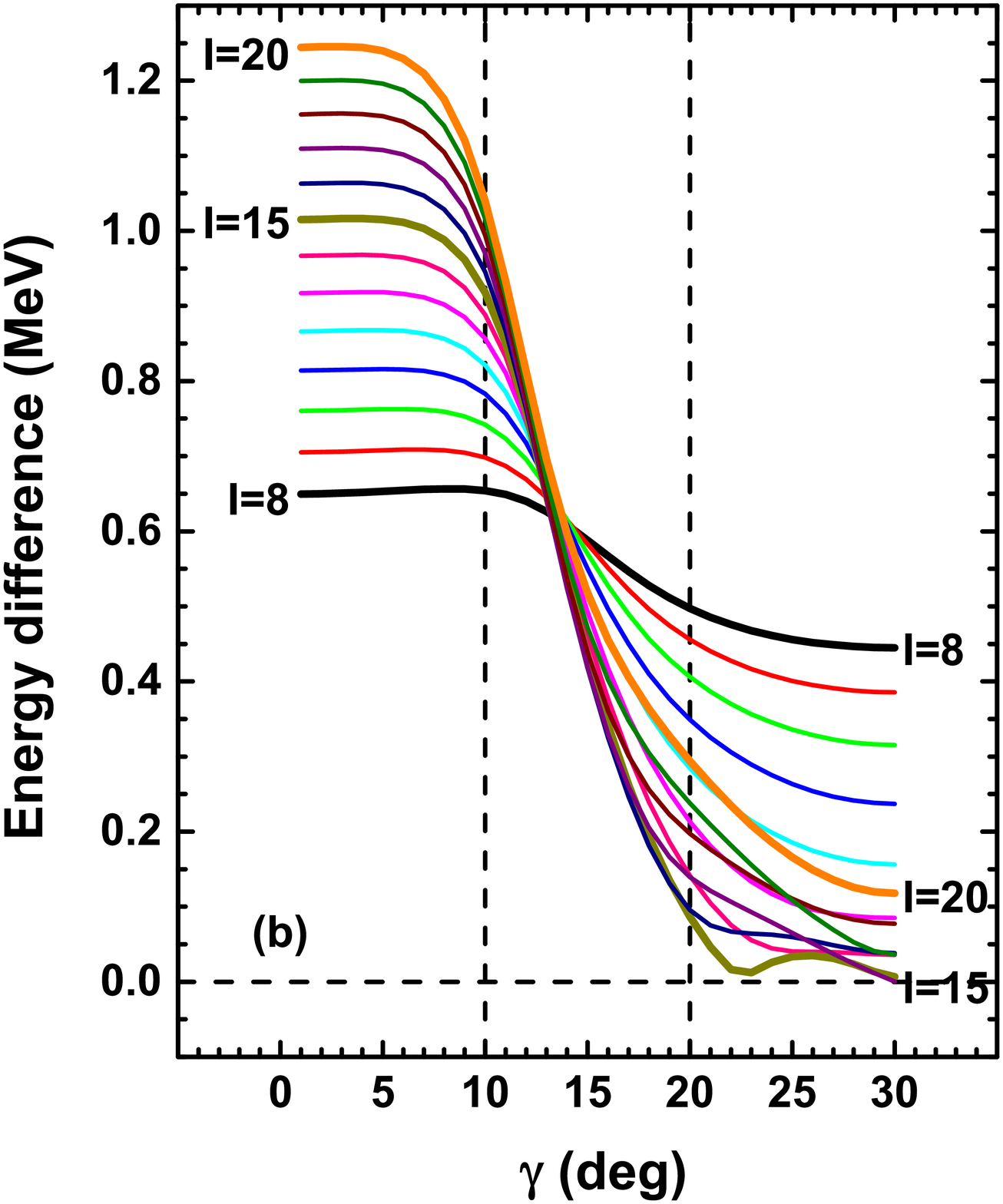}
    \caption{(a) Energy spectra of the yrast and yrare bands for the
    $\pi(1h_{11/2})\otimes \nu(1h_{11/2})^{-1}$ configuration calculated
    in the PRM as a function of triaxial deformation parameter $\gamma$.
    (b) Energy differences between the yrare and yrast bands as a function
    of $\gamma$.}\label{fig01}
  \end{center}
\end{figure}

The calculated energy spectra of the yrast and yrare bands for the
configuration $\pi(1h_{11/2})\otimes \nu(1h_{11/2})^{-1}$ as well
as their energy differences are shown as a function of triaxial
deformation parameter $\gamma$ in Fig.~\ref{fig01}(a) and (b) in the
spin region $8\hbar \leq I\leq 20\hbar$.

For $\gamma\leq 10^\circ$, the energy spectra of both yrast and yrare
bands do not vary significantly. Correspondingly, their energy differences
at each spin are almost constant, in particular for $\gamma \leq 5^\circ$.

For $10^\circ \leq \gamma \leq 20^\circ$, the energy spectra of the yrast and
yrare bands are sensitive to $\gamma$ and show different behaviors.
For the yrast band, one sees a slightly decreasing behavior for spins $8\hbar
\leq I \leq 11\hbar$, and a increasing behavior for $I \geq 12\hbar$. In contrast
to this, the yrare band decreases in the entire spin region, showing a
stronger decrease at high spins. Such behaviors of the yrast and yrare bands
cause the doublet bands to come close together, and hence their energy differences
decrease dramatically. For example, the energy difference at $I=20\hbar$ decreases
from about $1.0$ MeV to about $0.3$ MeV if $\gamma$ is varied from $10^\circ$
to $20^\circ$.

For $20^\circ \leq \gamma \leq 30^\circ$, only a slight variation of the
energy spectra of the yrast and yrare bands is observed. This feature
narrows further the energy gap between the doublet bands, making them more
degenerate. Actually, it becomes difficult to identify two separated rotational
bands in the spin region $12\hbar \leq I \leq 20\hbar$ when $\gamma$ reaches
$30^\circ$, since the energy differences are less than 200 keV. In many
publications~\cite{Frauendorf1997NPA, B.Qi2009PRC, Q.B.Chen2018PRC, Q.B.Chen2018PRC_v1},
$\gamma=30^\circ$ is considered as an ideal condition for the existence of
chiral rotation for the symmetric particle-hole configuration $\pi(1h_{11/2})
\otimes \nu(1h_{11/2})^{-1}$. Here, one observes from Fig.~\ref{fig01}(b)
that very good degeneracy and hence the condition for chiral rotation
occurs also at $\gamma \sim 23^\circ$ for $I=15\hbar$.

\subsection{Angular momenta}

\begin{figure*}[!ht]
  \begin{center}
    \includegraphics[width=7.5 cm]{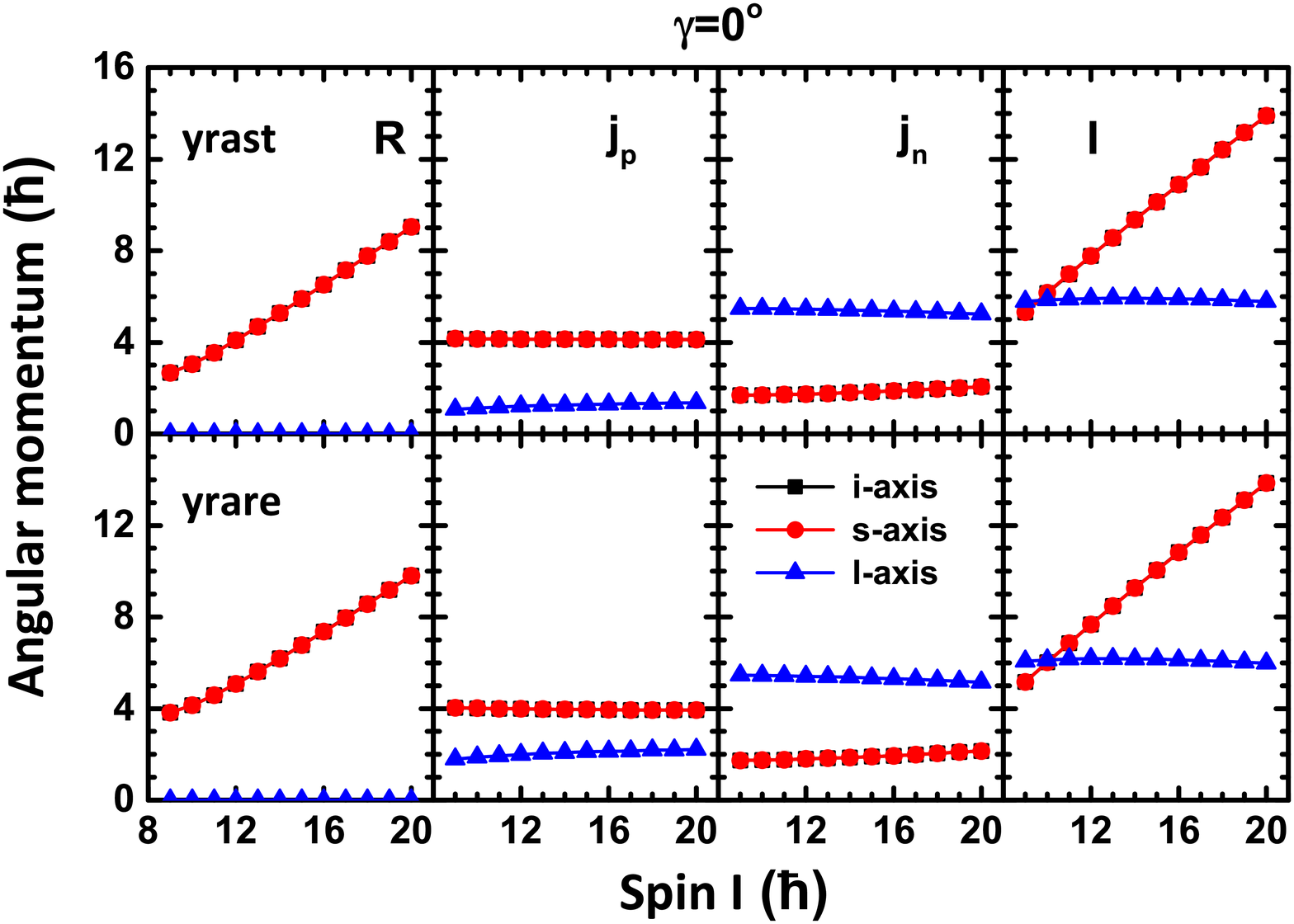}\quad
    \includegraphics[width=7.5 cm]{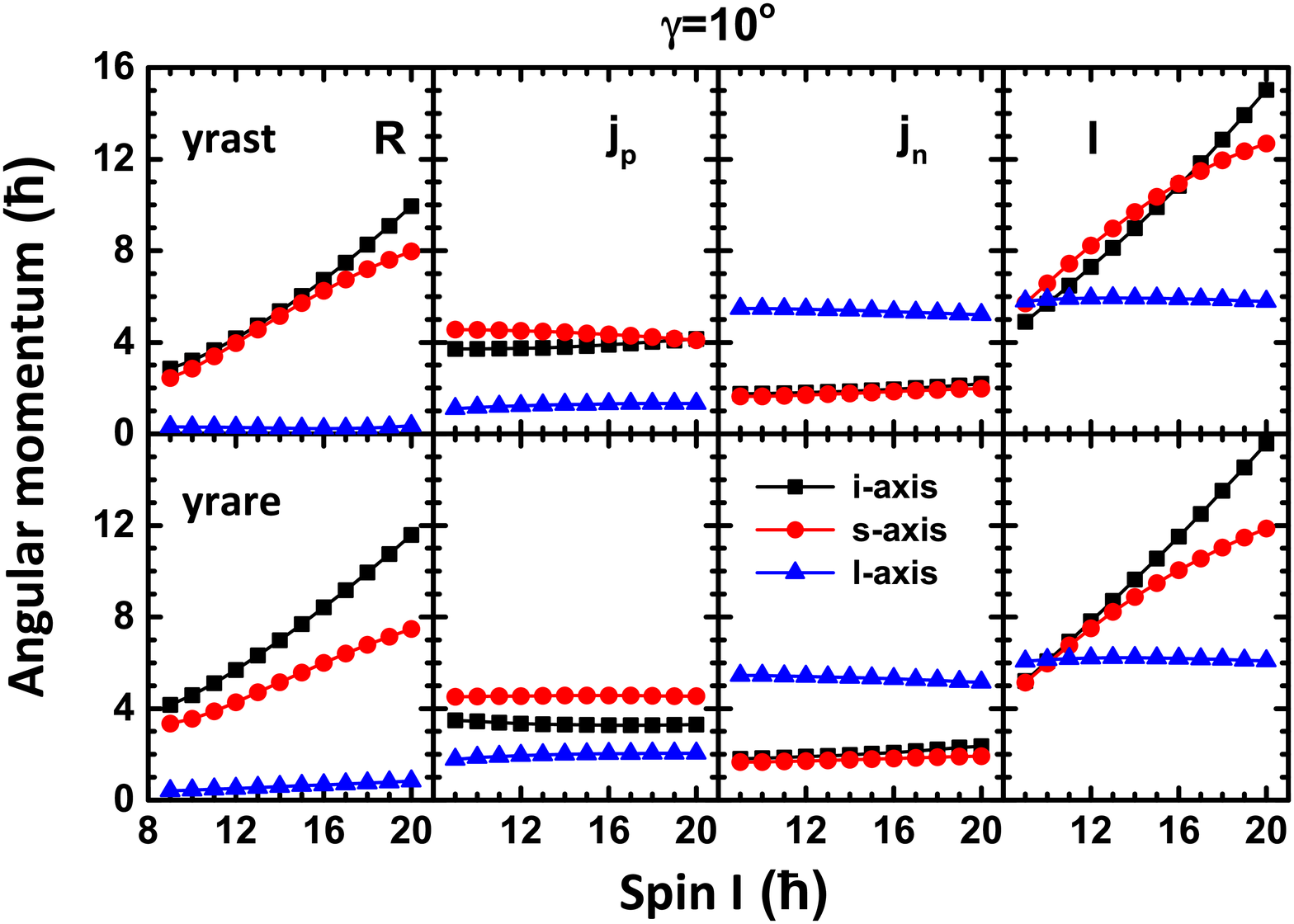}\\
    ~~\\
    \includegraphics[width=7.5 cm]{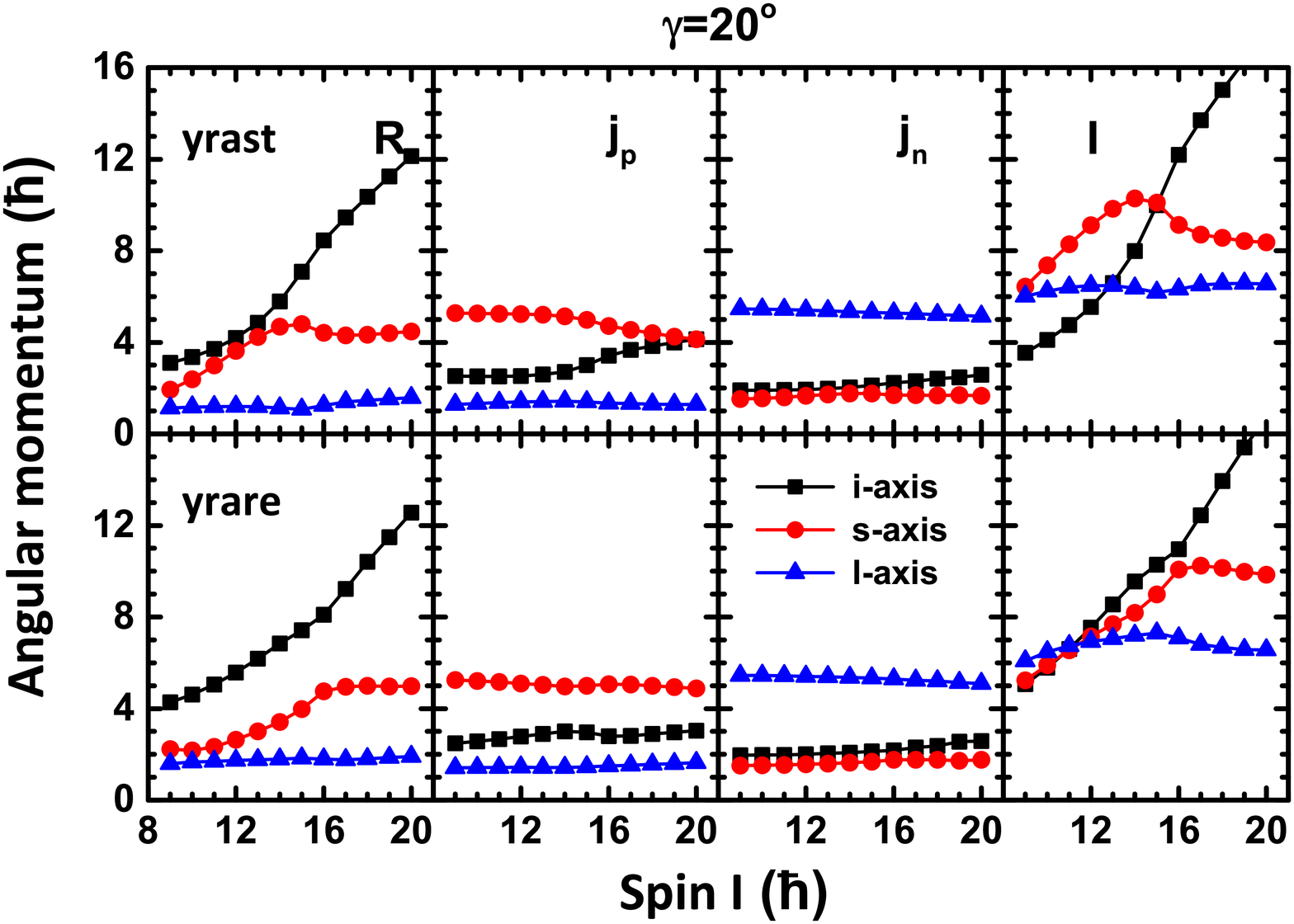}\quad
    \includegraphics[width=7.5 cm]{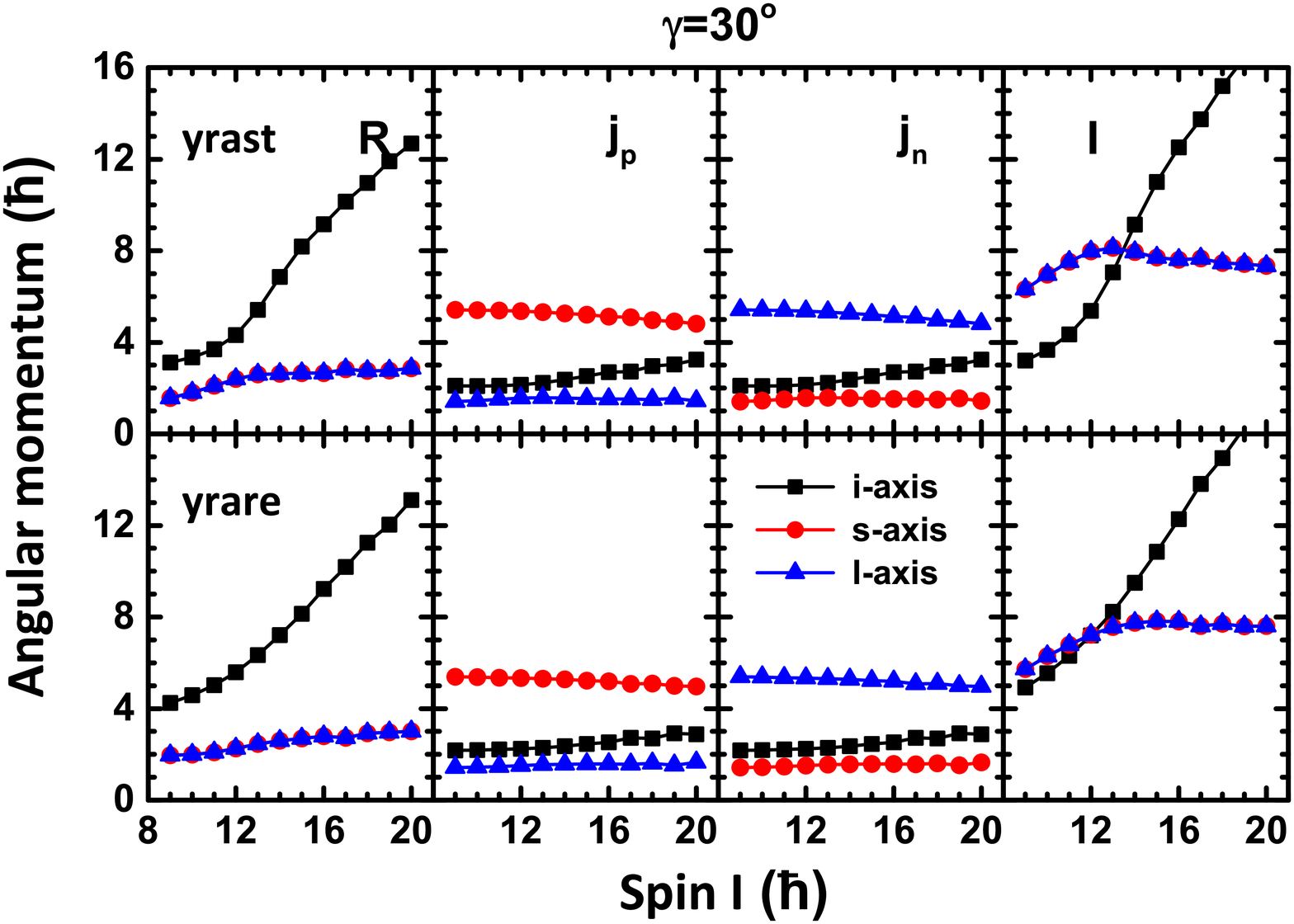}
    \caption{Root mean square values of angular momentum components along the
    intermediate ($i$-), short ($s$-), and long ($l$-) axes of the rotor, proton, neutron, and
    the total spin for the yrast and yrare bands calculated in the PRM at
    $\gamma=0^\circ$, $10^\circ$, $20^\circ$, and $30^\circ$.}\label{fig06}
  \end{center}
\end{figure*}

In the following, the angular momentum geometries of the doublet bands
are investigated by considering the situations at $\gamma=0^\circ$, $10^\circ$,
$20^\circ$, and $30^\circ$.

In Fig.~\ref{fig06}, the angular momentum components along the
intermediate ($i$-), short ($s$-), and long ($l$-) axes
of the rotor ($\bm{R}$), proton ($\bm{j}_p$), neutron ($\bm{j}_n$),
and the total spin ($\bm{I}$) for the yrast and yrare
bands calculated in the PRM are shown for $\gamma=0^\circ$,
$10^\circ$, $20^\circ$, and $30^\circ$.

For $\gamma=0^\circ$, the deformation of the rotor is prolate. The
lengths of the $s$- and $i$- axes and the corresponding principal
moments of inertia are identical, while the moment of inertia
with respect to the $l$-axis vanishes. Therefore, the angular
momentum components along the $s$- and $i$- axes are identical, and
the collective rotation can not happen about the $l$-axis. This is
exhibited clearly in Fig.~\ref{fig06}. Note that due to
the axial symmetry of the prolate nuclear shape with respect
to the $l$-axis, the motion of the system is a planar rotation.
Both components of the rotor angular momentum increase linearly
with the spin $I$, whereas for the proton and neutron the angular
momentum components remain almost constant. The proton particle is
mainly aligned along the $s$-/$i$- axes, while the neutron
hole aligns along the $l$-axis. With these features, the components
of the total spin $\bm{I}$ along the $s$-/$i$-axis also increases
linearly, and the component along the $l$-axis stays constant.
Moreover, the components of the rotor angular momentum are different
in the yrast and yrare bands. This behavior leads to the large energy
difference between the doublet bands, as shown in Fig.~\ref{fig01}.

When deviating from the prolate deformation, the nuclear shape becomes
slightly triaxial. The three principal axes of the ellipsoid have different
lengths, and to each corresponds a finite moment of inertia.
This makes collective rotations about any of the three axes possible.
For $\gamma=10^\circ$, the $l$-axis component of the rotor angular momentum
is small due to the small moment of inertia. For the rotor,
the components $R_s$ and $R_i$ are similar at the low spins in the yrast
band, but these two components are different from the bandhead upward
in the yrare band. For the proton, the components $j_{p s}$ and $j_{p i}$
are similar, and the $l$-axis component is again small. For the neutron,
the component $j_{n l}$ is larger than $j_{\nu s}$ and $j_{\nu i}$, which
are both similar. With these properties, the $s$- and $i$- components of
the total spin come out similar, and they are larger than the $l$-component.
One observes that due to the slight deviation from prolate deformation, the
angular momentum geometry at $\gamma=10^\circ$ does not change much in comparison
to that at $\gamma=0^\circ$. This explains why the energy spectra for $\gamma\leq 10^\circ$
do not vary significantly, as shown in Fig.~\ref{fig01}.

For $\gamma=20^\circ$, the three angular momentum components are different
for both the yrast and yrare bands. As the total spin $I$ increases, the components of
$\bm{R}$ increases gradually, while $\bm{j}_p$ and $\bm{j}_n$ move gradually towards
the $i$-axis. Hence, the three angular momentum form together the geometry for
aplanar rotation. The difference of orientation in the yrast and yrare bands
appears to come mainly from for the rotor for $I\leq 14\hbar$, and from the proton
for $I\geq 17\hbar$. At $I=15$ and $16\hbar$, the orientations of the rotor,
proton, neutron, and the total angular momentum in the yrast and yrare bands
become similar, and therefore their energy differences become smallest.

For $\gamma=30^\circ$, the moments of inertia corresponding to the $s$- and $l$- axes
are identical, and therefore $R_s=R_l$. The rotor mainly aligns along the $i$-axis
due to the largest momentum of inertia. In addition, one finds $j_{p s}=j_{n l}$,
$j_{p l}=j_{n s}$, and $j_{p i}=j_{n i}$, which leads to $I_s=I_l$. Similar to
the case $\gamma=20^\circ$, the difference of orientation in yrast and yrare bands
at $\gamma=30^\circ$ occurs mainly at low spins $I\leq 13\hbar$. This corresponds
to the picture of chiral vibration~\cite{B.Qi2009PRC, Q.B.Chen2018PRC_v1}. At
$15\hbar \leq I \leq 17\hbar$, the orientation in the yrast and yrare bands
are similar, and the doublet bands become almost degenerate, which leads chiral
rotation~\cite{B.Qi2009PRC, Q.B.Chen2018PRC_v1}.

\subsection{$R$-plots}

\begin{figure*}[!ht]
  \begin{center}
    \includegraphics[width=6.5 cm]{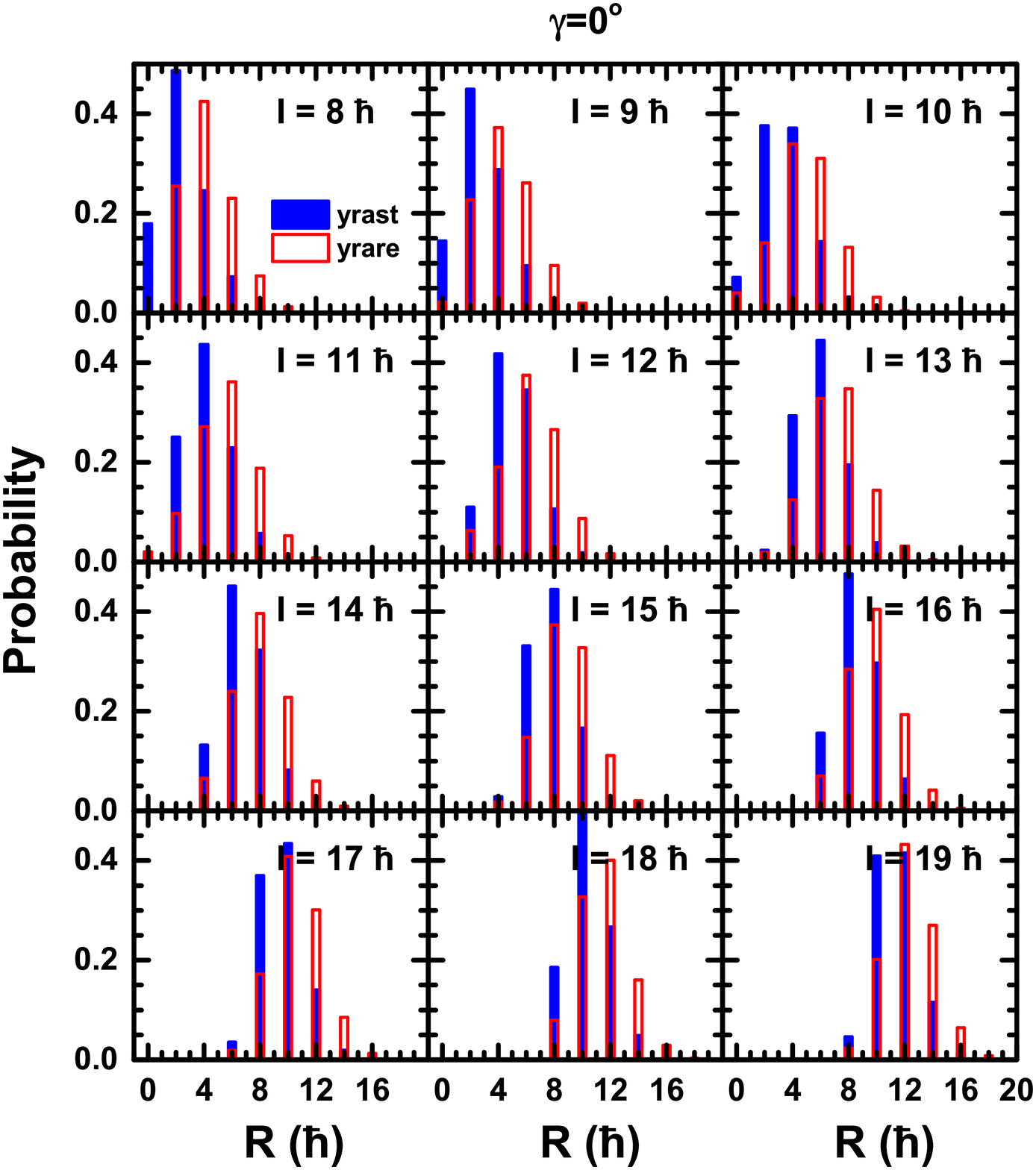}\quad
    \includegraphics[width=6.5 cm]{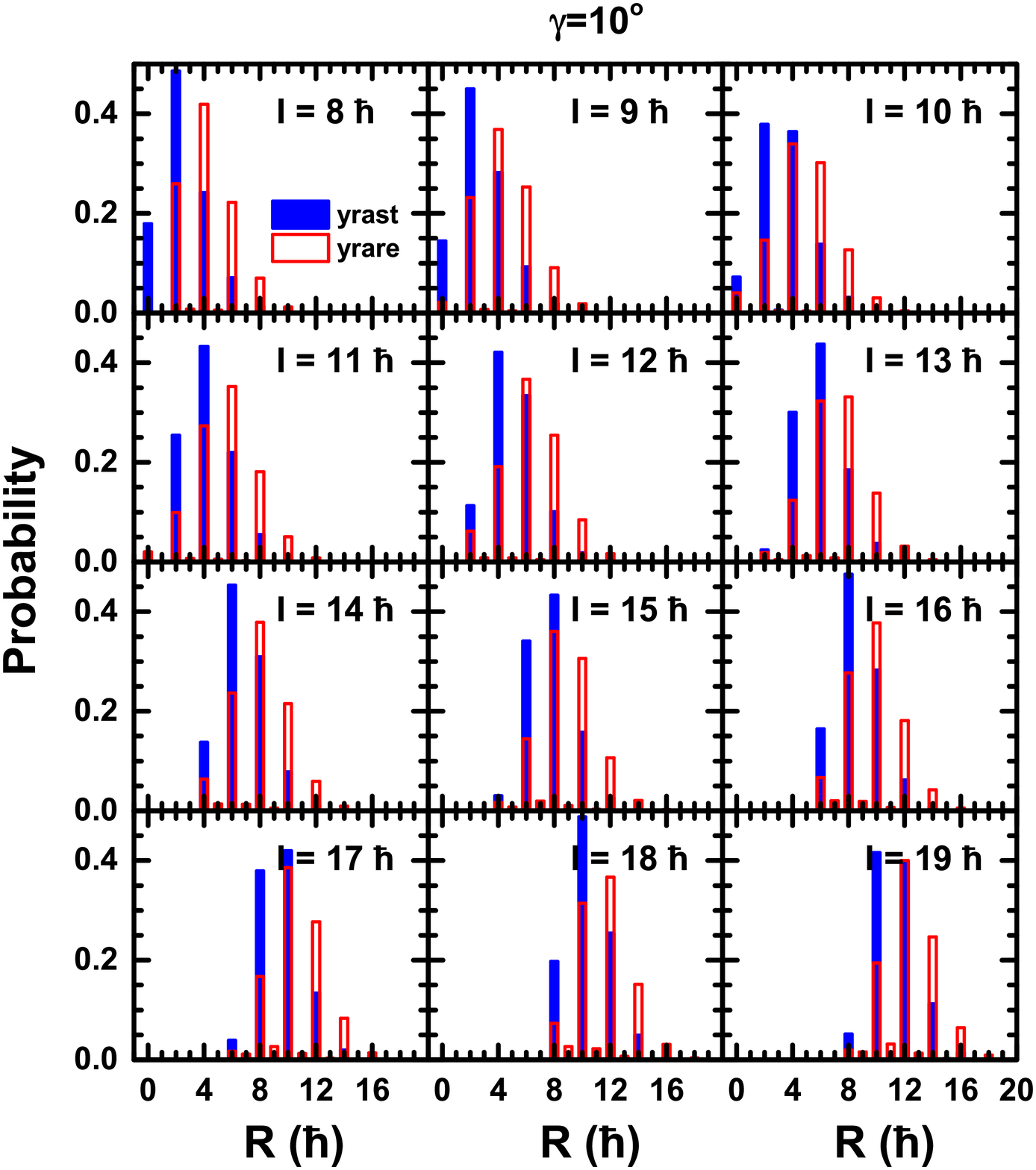}\\
    ~~\\
    \includegraphics[width=6.5 cm]{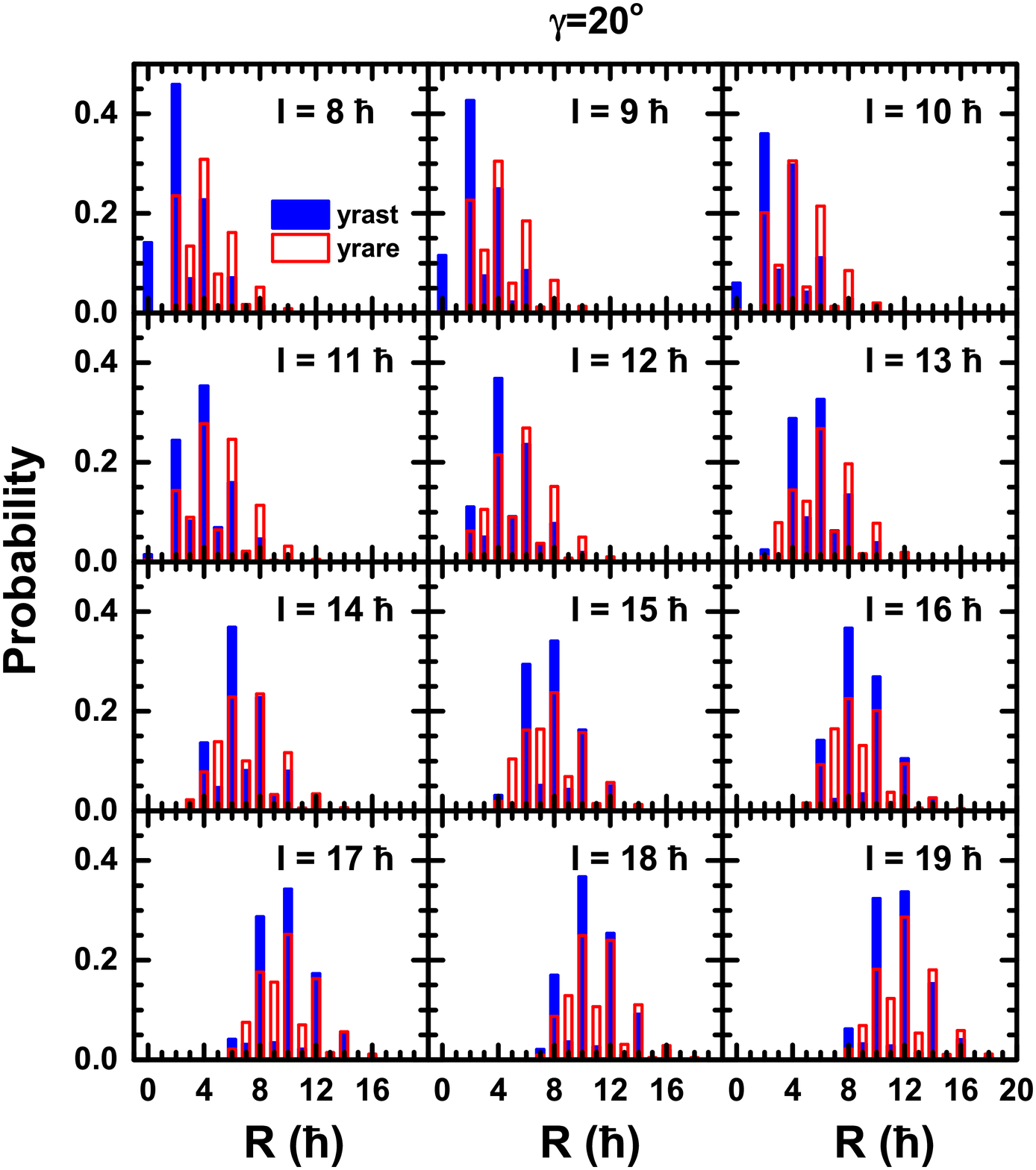}\quad
    \includegraphics[width=6.5 cm]{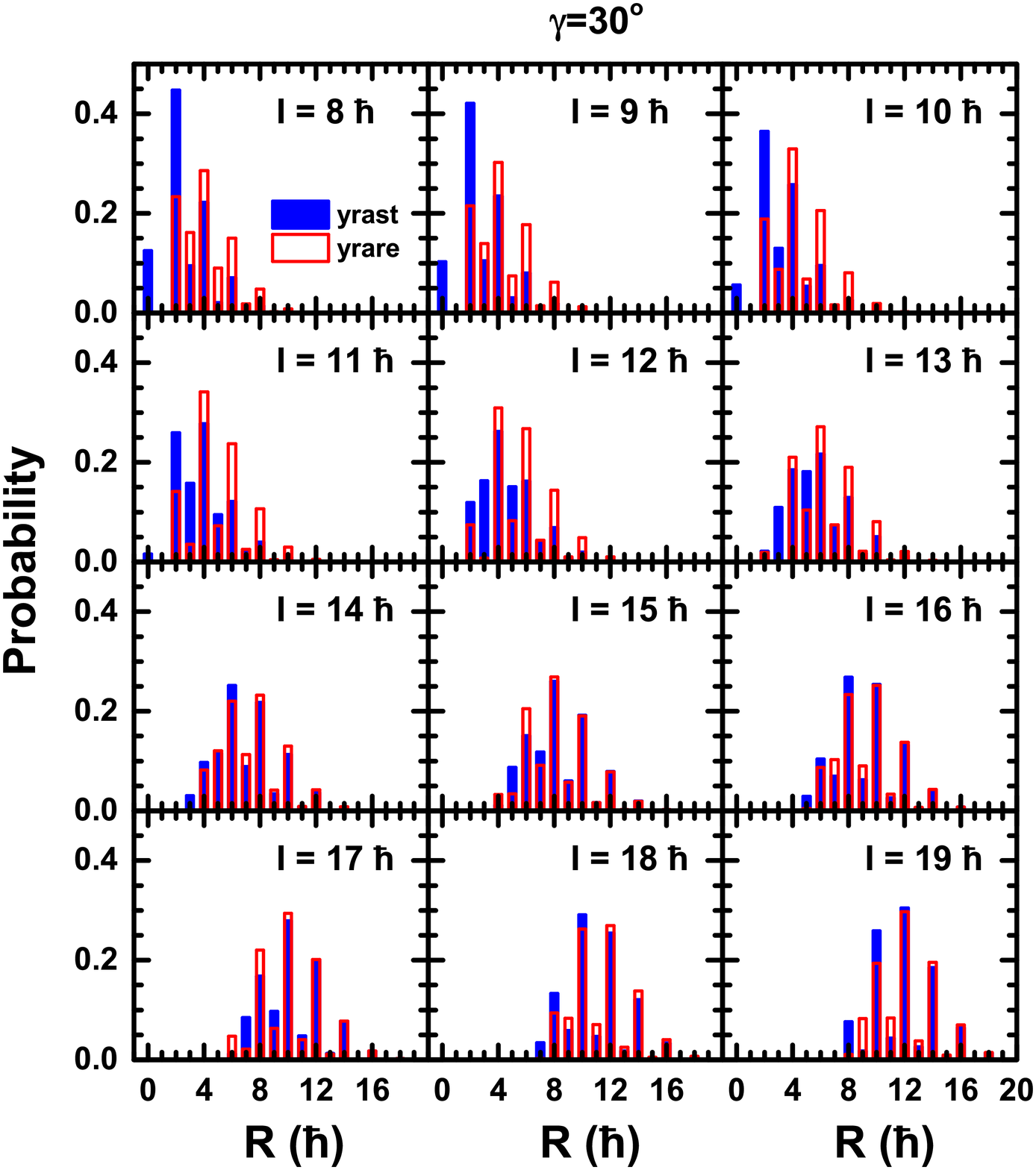}
    \caption{$R$-plots, probability distributions of the rotor angular momentum
    in the yrast and yrare bands for the $\pi(1h_{11/2})\otimes
    \nu(1h_{11/2})^{-1}$ configuration calculated with the PRM at $\gamma=0^\circ$,
    $10^\circ$, $20^\circ$, and $30^\circ$.}\label{fig02}
  \end{center}
\end{figure*}

In Fig.~\ref{fig02} the probability distributions $P_R$ of the rotor angular
momentum ($R$-plots) in the yrast and yrare bands for the $\pi(1h_{11/2})\otimes
\nu(1h_{11/2})^{-1}$ configuration calculated at $\gamma=0^\circ$,
$10^\circ$, $20^\circ$, and $30^\circ$ are shown. One observes that with
increasing total spin $I$, the distributions $P_R$ shift their weights
from the low $R$ to the high $R$ region, indicating a gradual increase
of the rotor angular momentum.

At $\gamma=0^\circ$ the quantum number $R$ can take only even integer values
since $K_R$ must be zero, and therefore the distribution $P_R$ is zero
at odd $R$. The $R$-plots for the yrast and yrare bands show a different
behavior in the whole spin region $8\hbar\leq I\leq 20\hbar$. The weights
at each $R$-value as well as the positions of the maxima are different.
In general, the $R$-value with maximal weight in the yrare band is
$2\hbar$ larger than that in the yrast band. Such a behavior causes
a large energy difference between the doublet bands.

At $\gamma=10^\circ$, the $R$-plot is quite similar to that at $\gamma=0^\circ$.
There are only some very small contributions at odd $R$-values in
the high-spin region.

At $\gamma=20^\circ$, the weights at odd $R$-values are more substantial.
This is due to the fact the energies of the rotor for odd-$R$ decrease
with increasing $\gamma$ and gradually become comparable to those
for even-$R$ at $\gamma=20^\circ$~\cite{Davydov1958NP}. For $I\leq 12\hbar$,
the $R$-value with maximal weight in the yrare band is still
$2\hbar$ larger than that in the yrast band. For $I \geq 13\hbar$,
the $R$-plots for the yrast and yrare bands are quite similar,
although the detailed amplitudes are a bit different. This similarity
leads to the small energy differences (less than 300 keV) for the
doublet bands in this spin region.

At $\gamma=30^\circ$, the most prominent feature is that the $R$-plots of the
yrast and yrare bands are very similar for $I\geq 14\hbar$, concerning the
distribution patterns and also the amplitudes. These properties lead to
the degenerated doublet bands.

\subsection{$K_R$-plots}

\begin{figure*}[!ht]
  \begin{center}
    \includegraphics[width=6.5 cm]{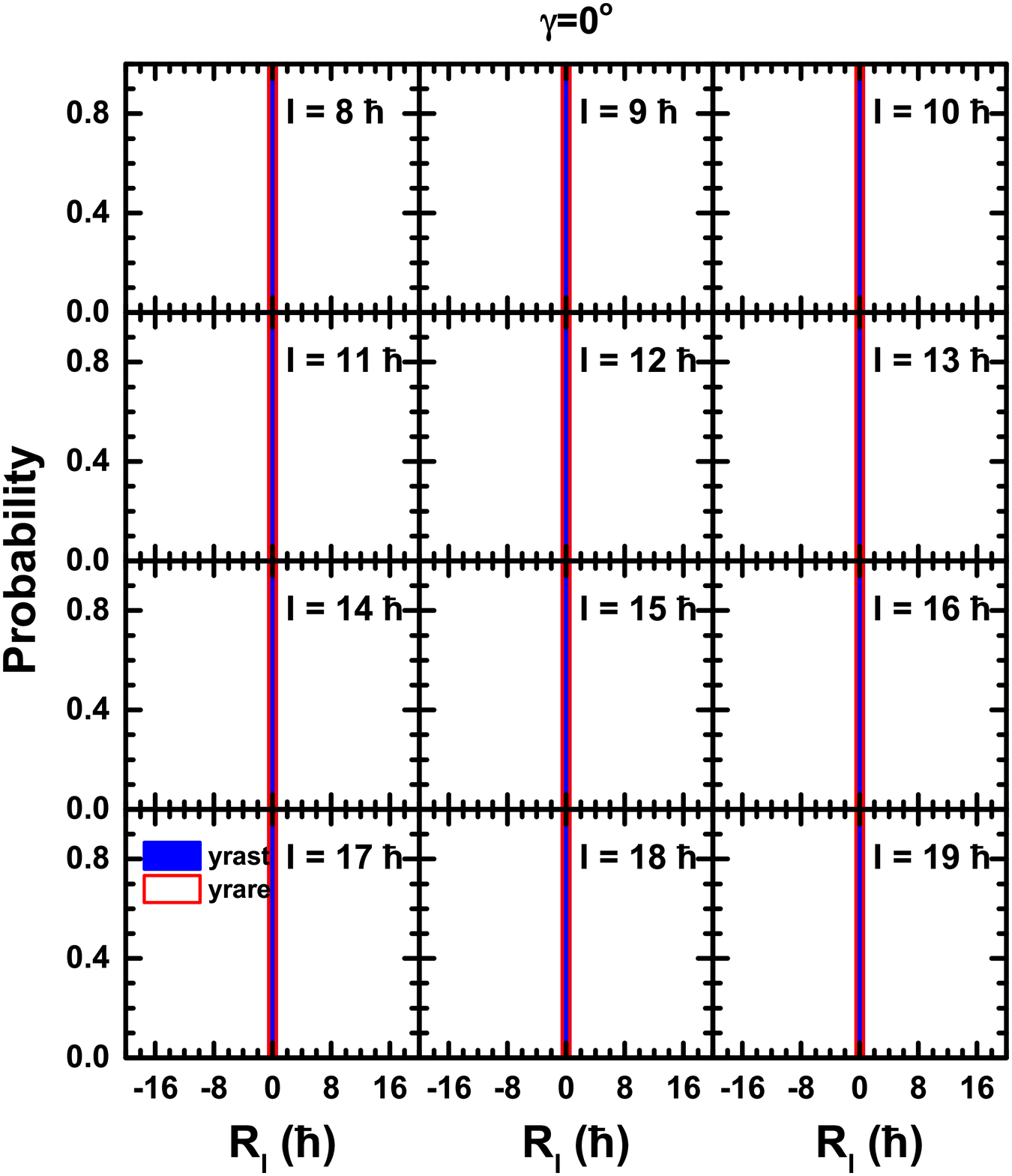}\quad
    \includegraphics[width=6.5 cm]{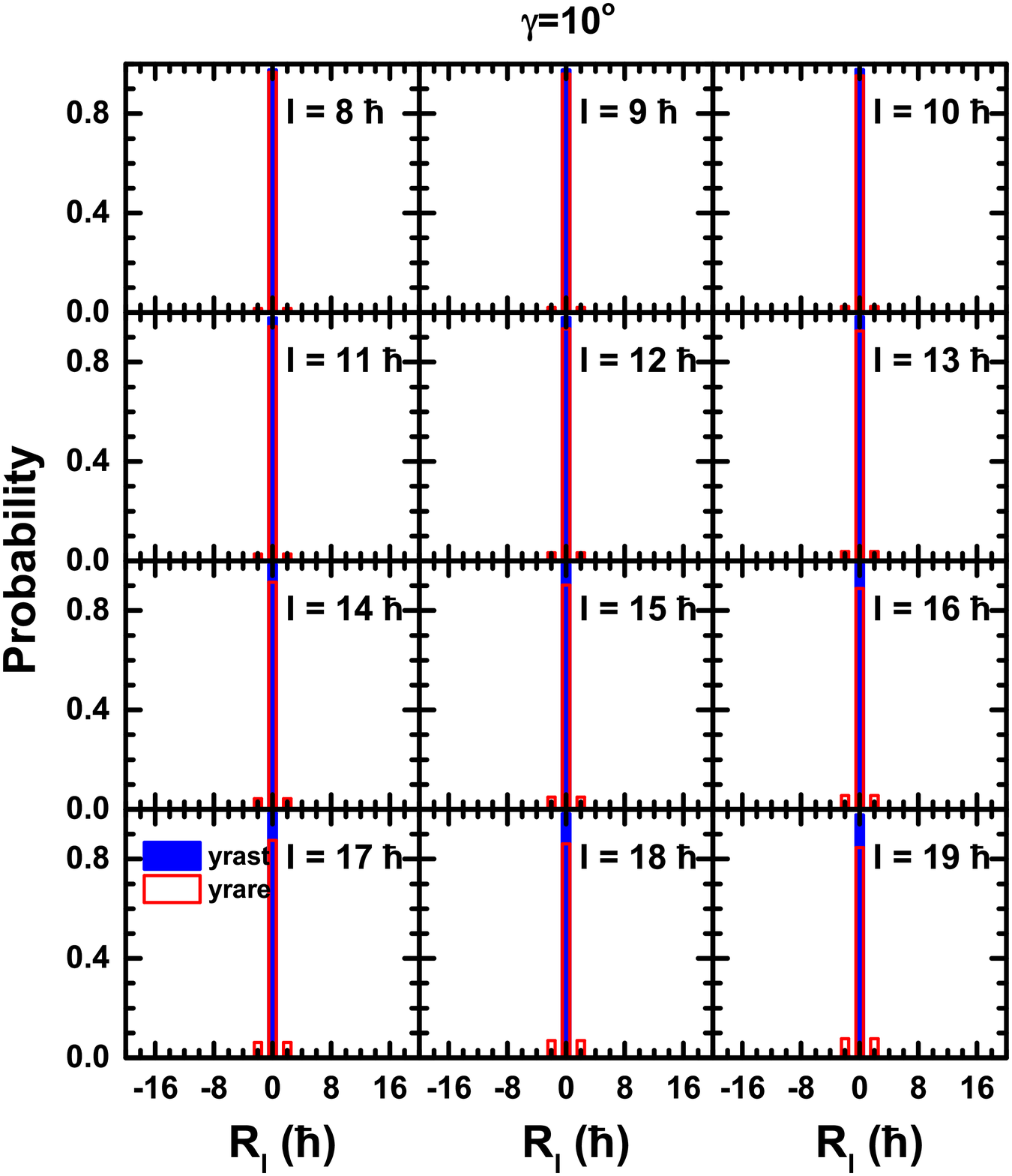}\\
    ~~\\
    \includegraphics[width=6.5 cm]{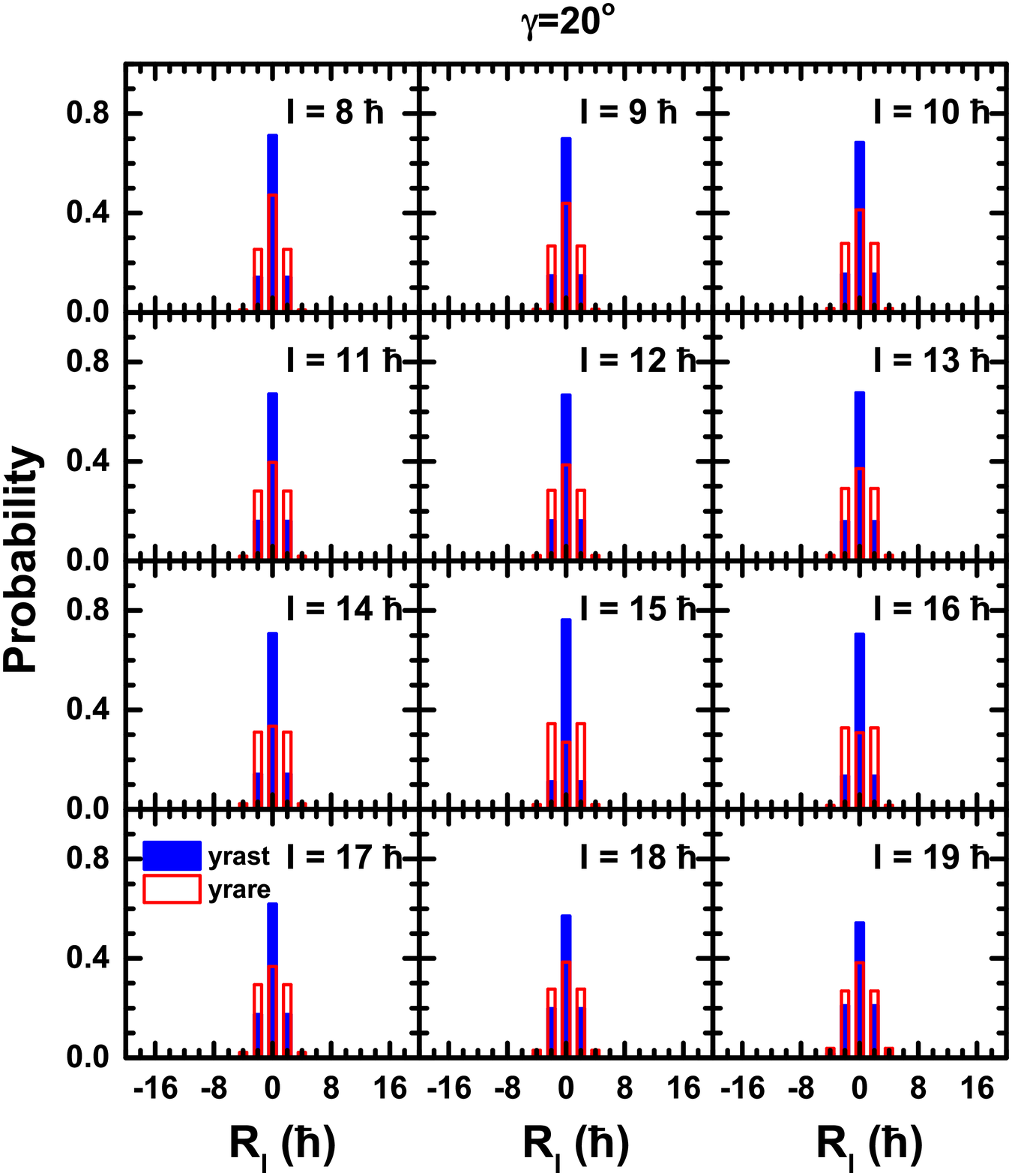}\quad
    \includegraphics[width=6.5 cm]{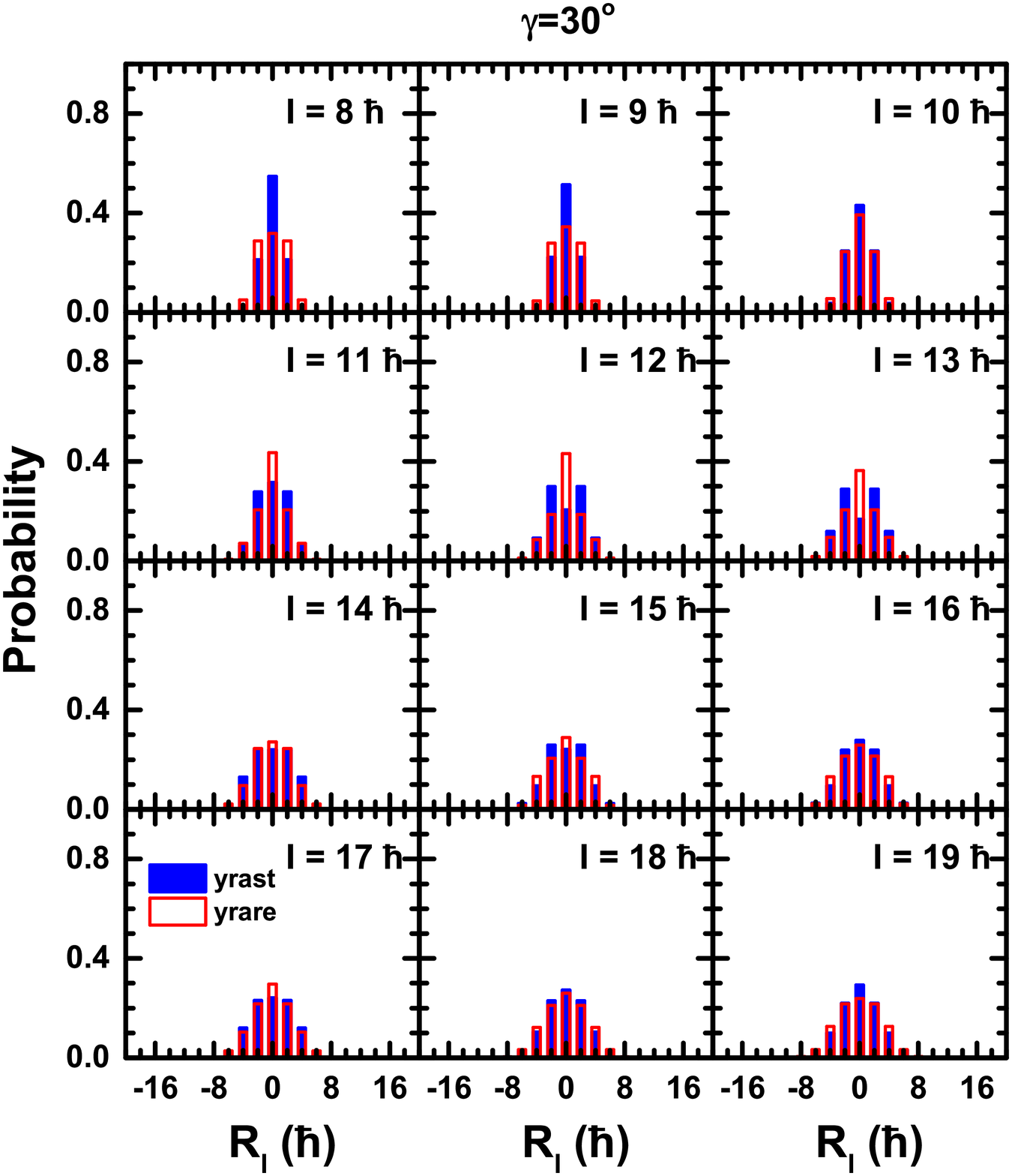}
    \caption{$K_{R_l}$-plots, probability distributions for the projection of the
    rotor angular momentum onto the $l$-axis in the yrast and yrare bands for the
    $\pi(1h_{11/2})\otimes \nu(1h_{11/2})^{-1}$ configuration calculated
    with the PRM at $\gamma=0^\circ$, $10^\circ$, $20^\circ$, and $30^\circ$.}\label{fig03}
  \end{center}
\end{figure*}

In the following the probability distributions for the projections ($K_R=R_l$, $R_s$,
and $R_i$) of the rotor angular momentum onto the $l$-, $s$-, and $i$-axes ($K_R$-plots)
will be investigated. For $\gamma \in [0^\circ, 30^\circ]$, the $l$-axis is the designated
quantization axis. The distributions with respect to the $s$- and $i$-axis are obtained by
taking $\gamma+120^\circ$ and $\gamma+240^\circ$. These $\gamma$-values correspond
to the equivalent sectors such that the nuclear shape remains the same, and only
the principal axes get interchanged~\cite{Bohr1975, Ring1980book}. The
$K_R$-plots are symmetric under $K_R \to -K_R$ due to the $D_2$
symmetry of the triaxial nucleus.

In Fig.~\ref{fig03}, the probability distributions for the projection
of the rotor angular momentum onto the $l$-axis $P_{R_l}$ are shown
for the yrast and yrare bands at $\gamma=0^\circ$, $10^\circ$, $20^\circ$,
and $30^\circ$.

At $\gamma=0^\circ$, $l$-axis component of the rotor angular momentum
vanishes. Hence, $P_{R_l}=1$ at $R_l=0$ for all spin states.

At $\gamma=10^\circ$, the $K_{R_l}$-plots are similar to those at $\gamma=0^\circ$.
There appear only some small distributions at $R_l=\pm 2\hbar$ in the high-spin
region of the yrare band. This is consistent with the picture
that the $l$-axis component of rotor angular momentum is small.

At $\gamma=20^\circ$, the $K_{R_l}$ distributions have weights mainly at
$R_l=0$, $\pm 2\hbar$, indicating still small $l$-axis component of the
rotor angular momentum.

For $\gamma=30^\circ$, the $K_{R_l}$ distribution becomes wider, and one observes
non-vanishing contributions at $R_l=\pm 4\hbar$. Moreover, for $I\geq 14\hbar$,
the distributions in the yrast and yrare bands are quite similar.

\begin{figure*}[!ht]
  \begin{center}
    \includegraphics[width=6.5 cm]{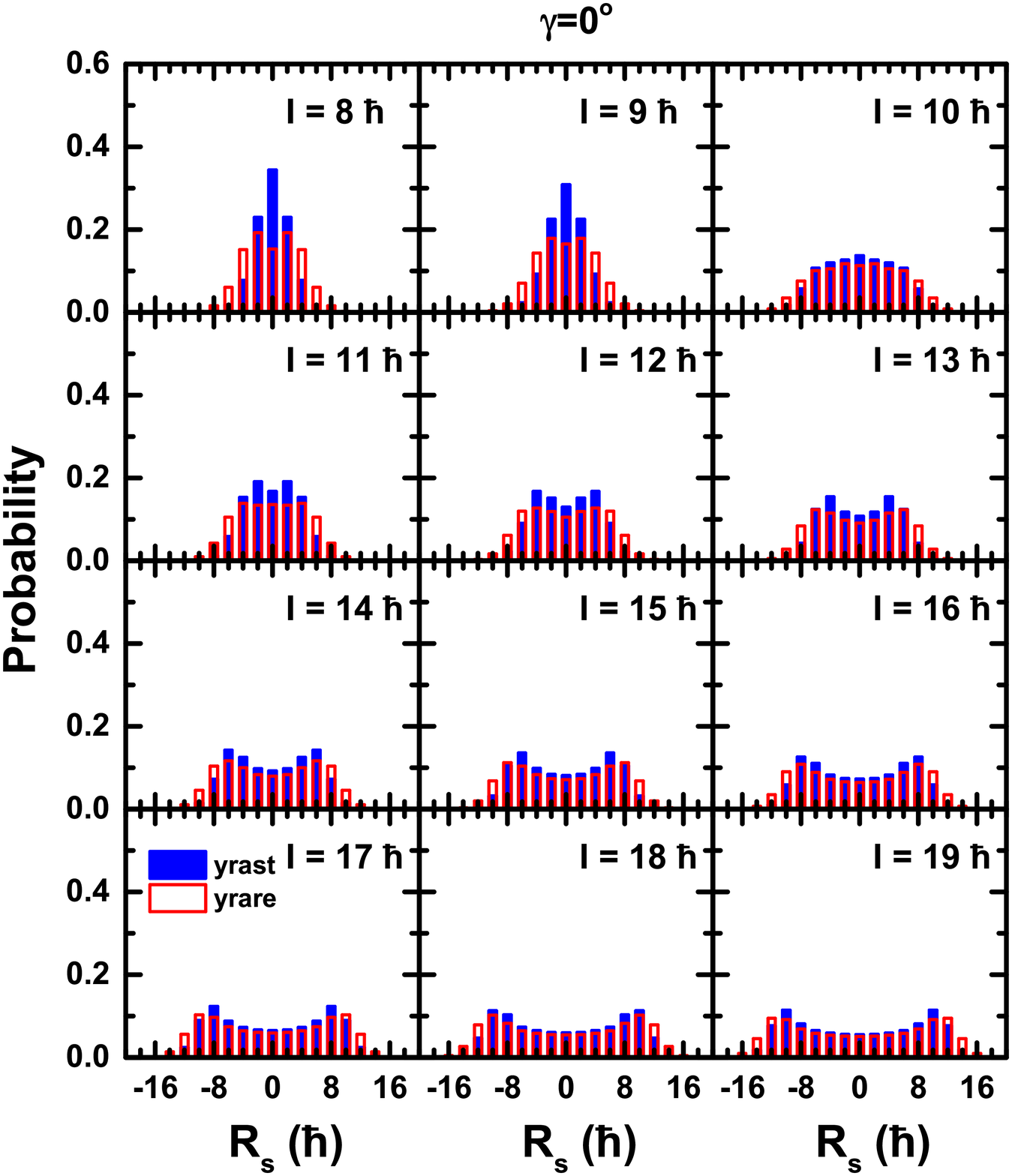}\quad
    \includegraphics[width=6.5 cm]{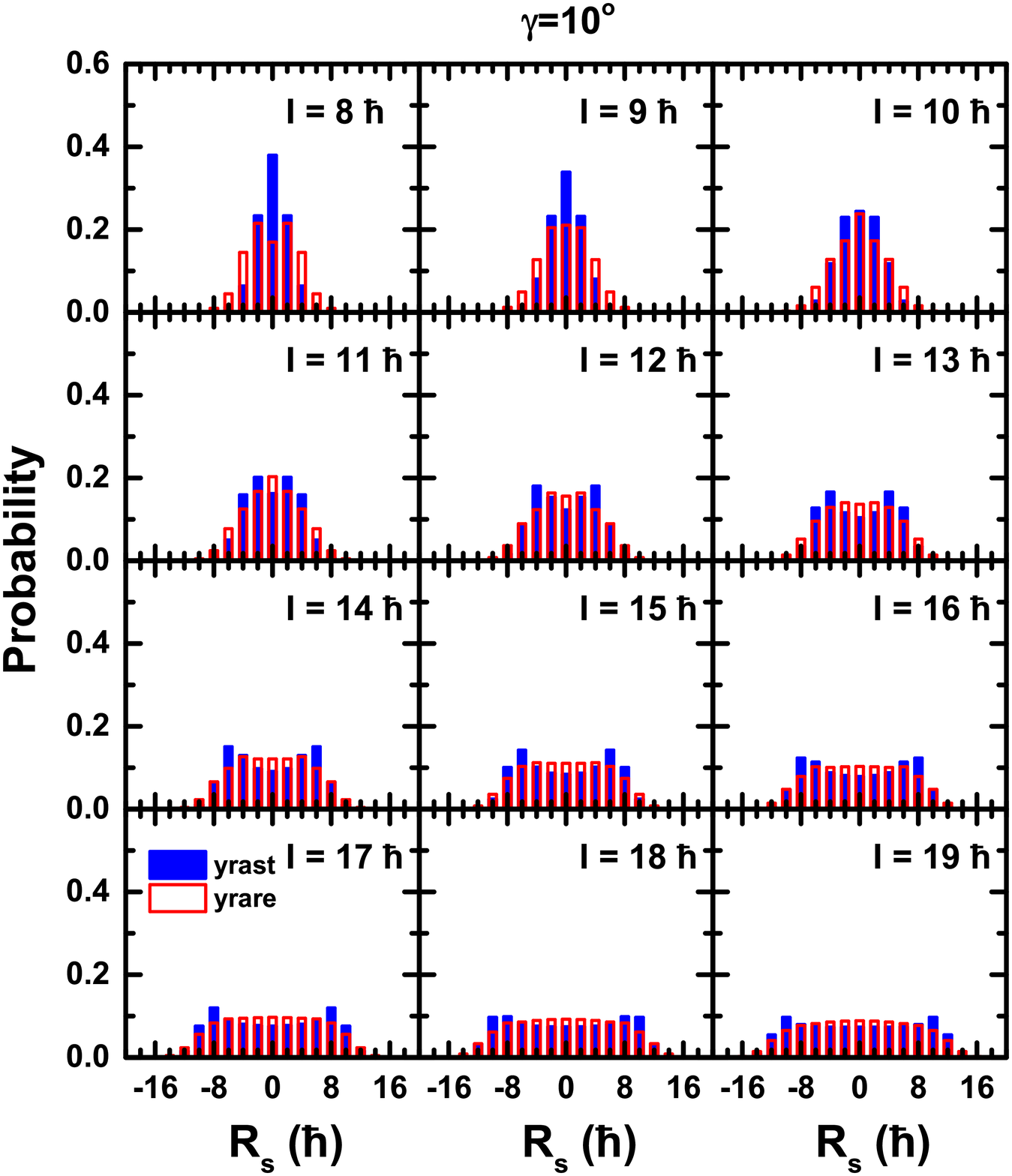}\\
    ~~\\
    \includegraphics[width=6.5 cm]{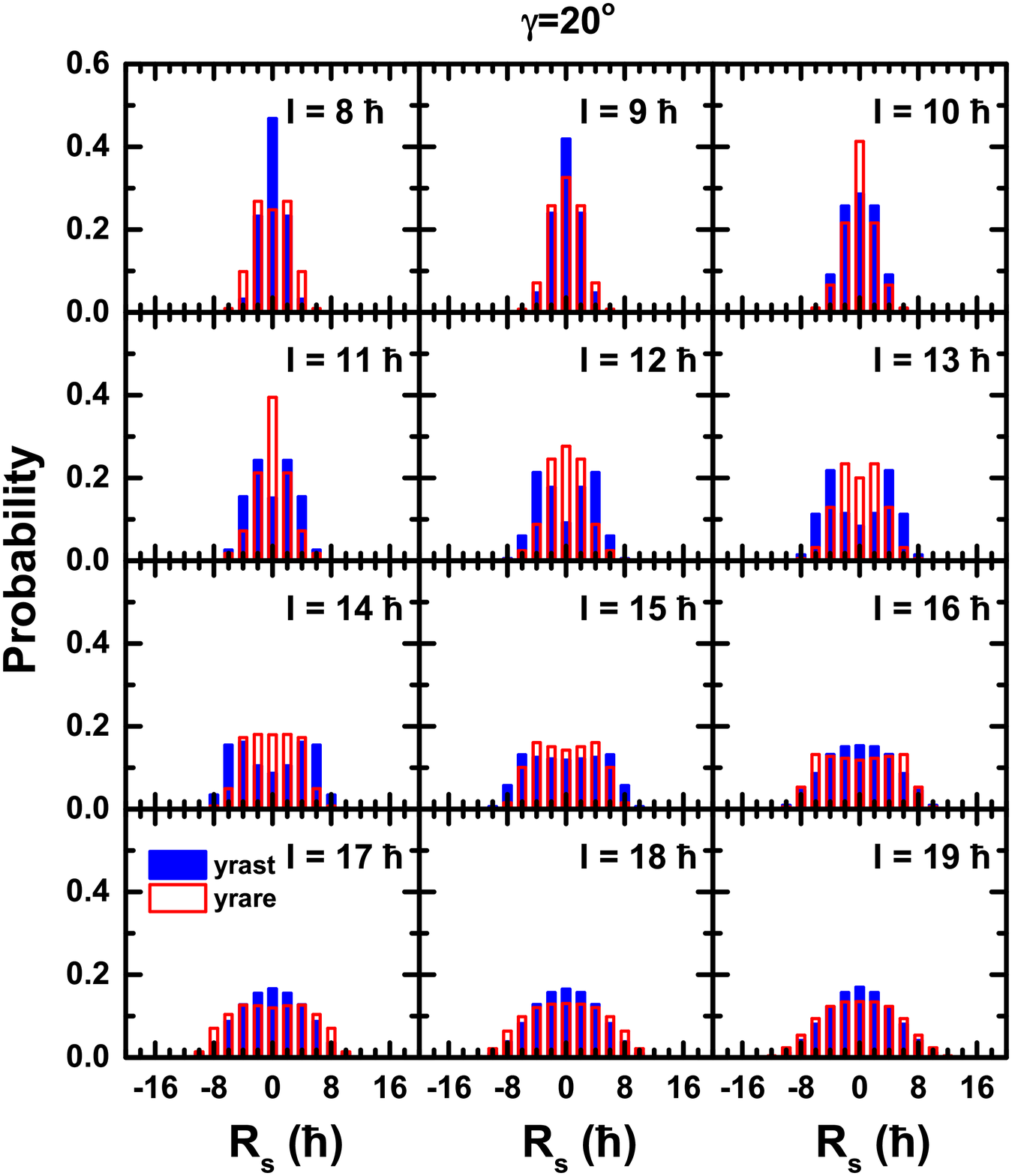}\quad
    \includegraphics[width=6.5 cm]{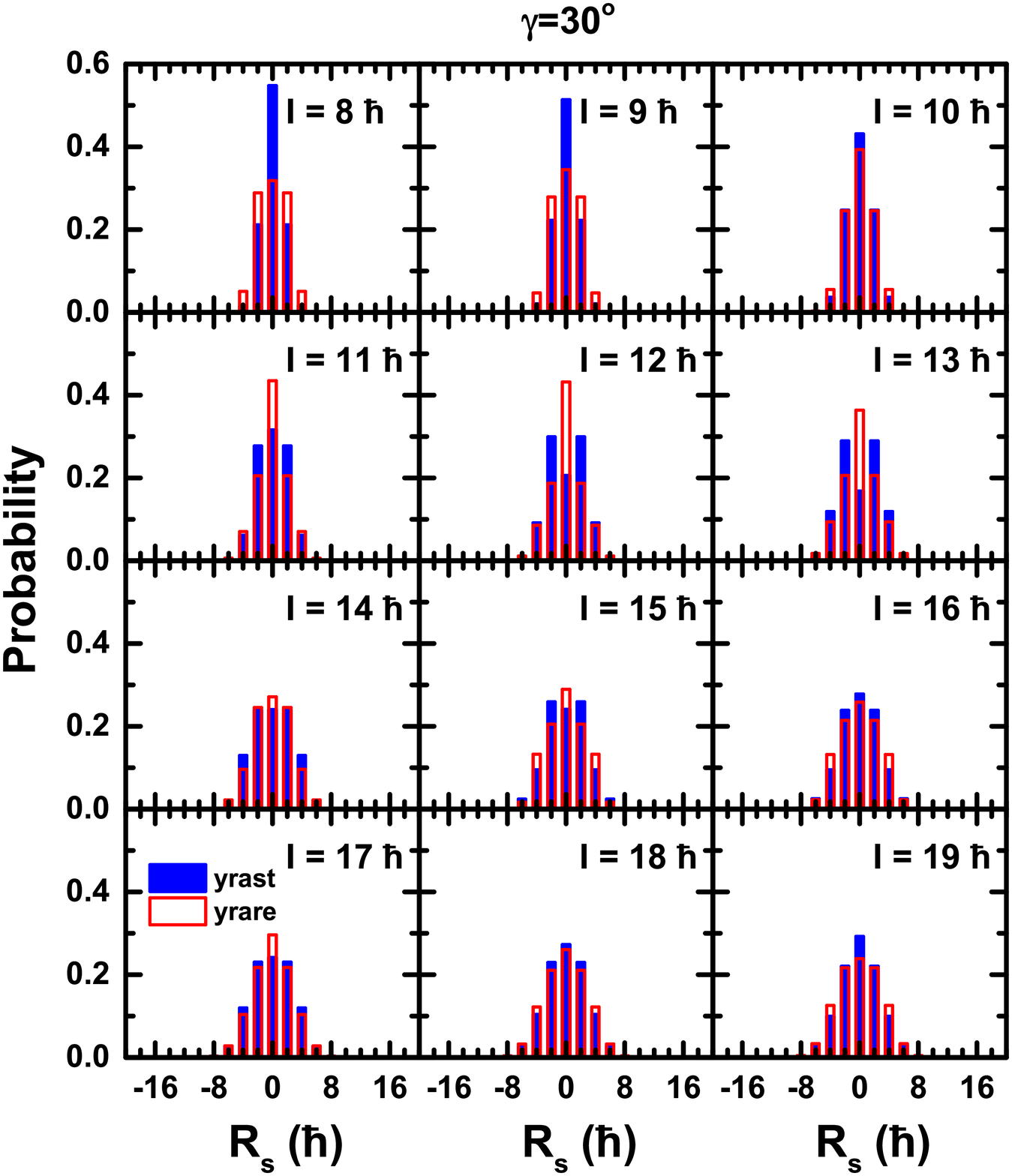}
    \caption{Same as Fig.~\ref{fig03}, but for the projection onto the $s$-axis.}\label{fig04}
  \end{center}
\end{figure*}

The probability distributions $P_{R_s}$ of the component $R_s$ are displayed
in Fig.~\ref{fig04} for the yrast and yrare bands at $\gamma=0^\circ$,
$10^\circ$, $20^\circ$, and $30^\circ$.

For $\gamma=0^\circ$, the distribution $P_{R_s}$ has a wide spread with
a peak around  $R_s=0\hbar$ at low spins $I=8$ and $9\hbar$. For $I\geq 10\hbar$,
this peak moves gradually towards large $R_s$ value, indicating the
increase of the rotor angular momentum component along $s$-axis.

At $\gamma=10^\circ$, the $P_{R_s}$ distributions in the doublet bands have
again a peak around $R_s=0\hbar$ for low spins $I=8$ and $9\hbar$.
For $I\geq 10\hbar$, the $P_{R_s}$-plots of the doublet bands behave differently.
In the yrast band it has two distinct peaks located at nonzero $R_s$,
whereas in the yrare band it is rather broad with a peak
at $R_s=0\hbar$. This implies a larger mean square deviation
$\langle R_s^2 \rangle$ in the yrast band compared to yrare band.

At $\gamma=20^\circ$, the $P_{R_s}$ distributions show a more complicated
behavior with increasing spin. For $I\leq 11\hbar$, one finds peaks
around $R_s=0\hbar$ for both yrast and yrare bands. In the region $12\hbar
\leq I \leq 15\hbar$, the peak in the yrast band occurs at nonzero $R_s$-values,
while in the yrare band it stays at $R_s=0\hbar$. For $I\geq 16\hbar$, the
$R_s$-plots of the yrast and yrare bands are again similar with a peak at
$R_s=0\hbar$.

At $\gamma=30^\circ$, the $P_{R_s}$ and $P_{R_l}$ distributions are the same
since moments of inertia with respect to $l$- and $s$-axes are identical.
The $P_{R_s}$ distributions becomes narrow in comparison to the other cases of
triaxial deformation. The peaks located around at $R_s=0\hbar$ demonstrate
the reduction of the rotor angular momentum component along the $s$-axis,
as shown by the four plots in Fig.~\ref{fig06}.

\begin{figure*}[!ht]
  \begin{center}
    \includegraphics[width=6.5 cm]{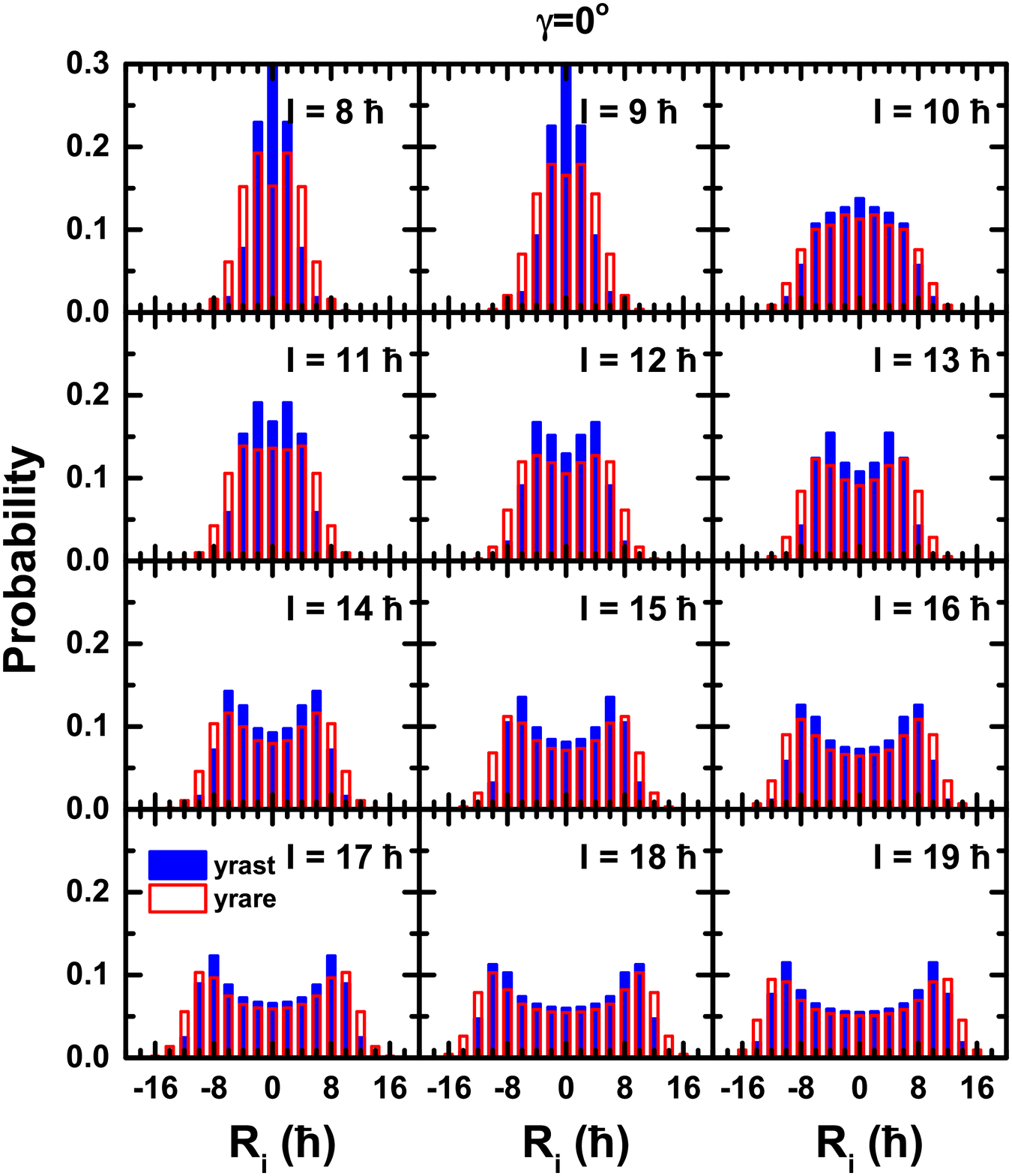}\quad
    \includegraphics[width=6.5 cm]{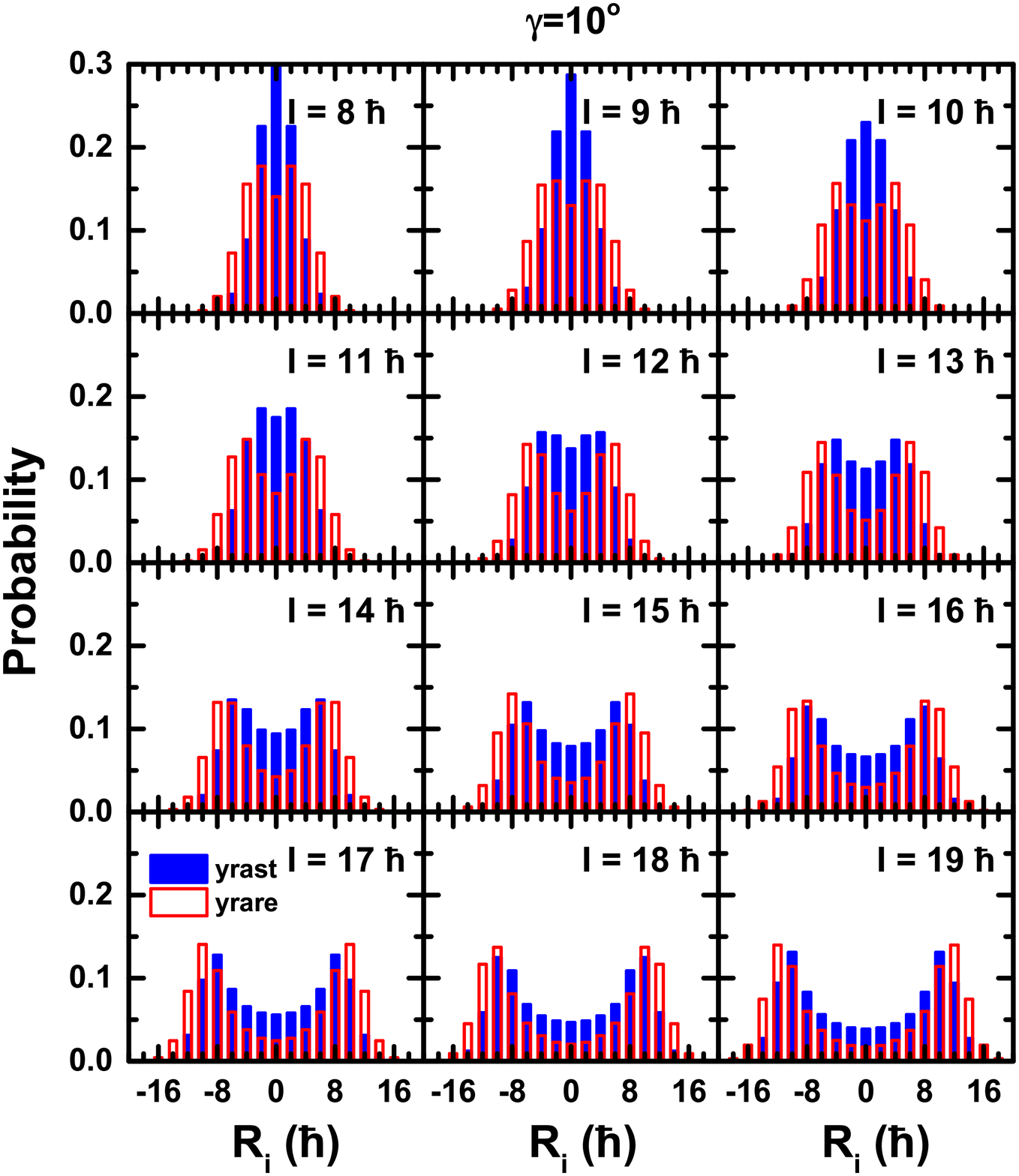}\\
    ~~\\
    \includegraphics[width=6.5 cm]{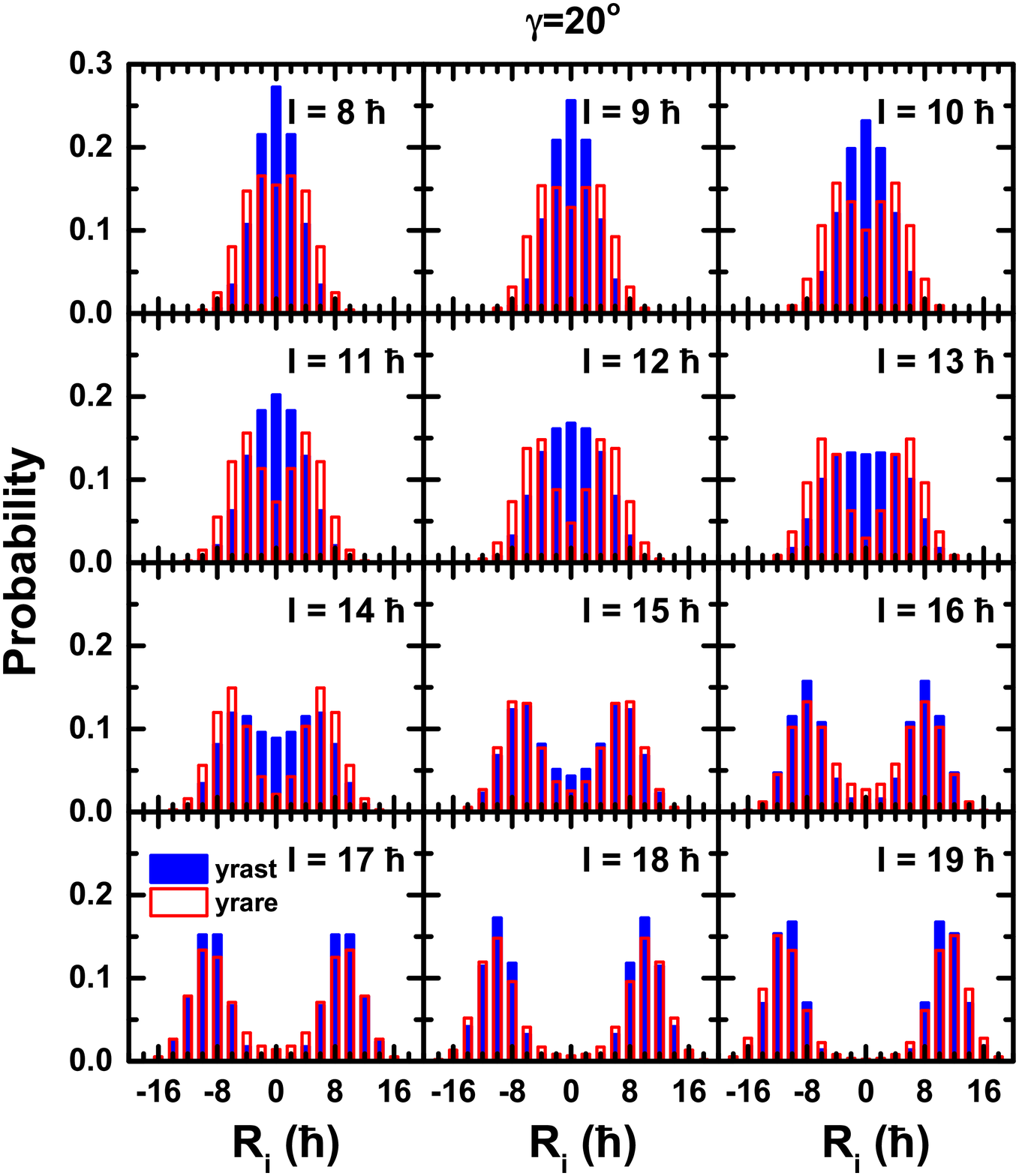}\quad
    \includegraphics[width=6.5 cm]{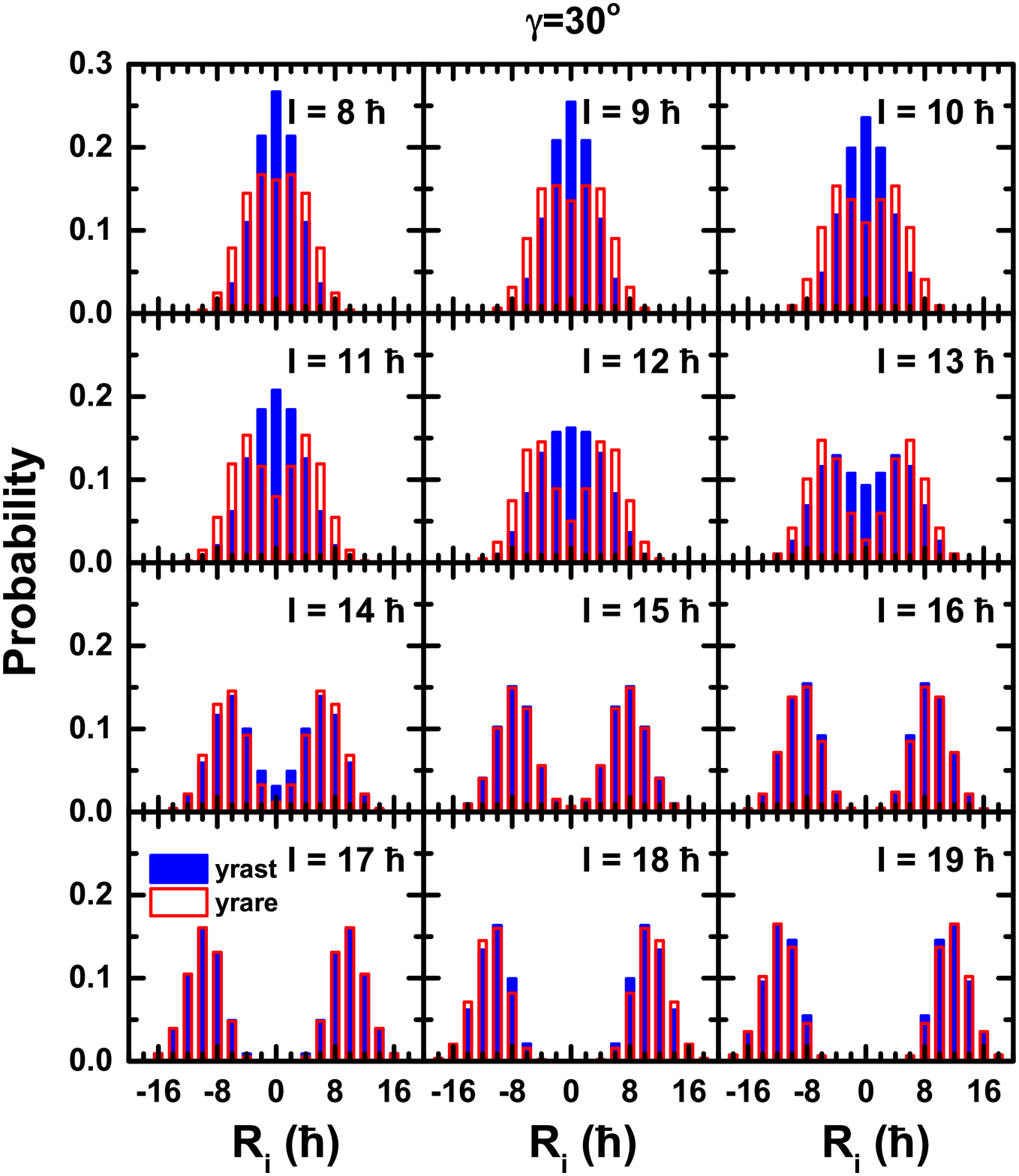}
    \caption{Same as Fig.~\ref{fig03}, but for the projection onto the $i$-axis.}\label{fig05}
  \end{center}
\end{figure*}

In Fig.~\ref{fig05} the probability distributions $P_{R_i}$ of the component
$R_i$ are shown for the yrast and yrare bands at $\gamma=0^\circ$, $10^\circ$,
$20^\circ$, and $30^\circ$.

At $\gamma=0^\circ$, the $P_{R_i}$ and $P_{R_s}$ distributions are the same
since the corresponding moments of inertia are equal.

At $\gamma=10^\circ$ the $P_{R_i}$ distribution is similar to that at
$\gamma=0^\circ$, but the amplitude at $R_i=0\hbar$ is a bit larger in the
yrast band than in the yrare band.

At $\gamma=20^\circ$, the $P_{R_i}$ distribution for $I\leq 13\hbar$ in the yrast band
has only one peak at $R_i=0\hbar$, while that in the yrare band has two peaks at
nonzero $R_i$. This situation corresponds to the chiral vibration. For $I\geq 14\hbar$,
the $P_{R_i}$ distributions for both doublet bands have two peaks at nonzero $R_i$,
indicating the onset of chiral rotation.

At $\gamma=30^\circ$, the behavior of the $P_{R_i}$ distribution is similar to that at
$\gamma=20^\circ$. It shows the picture of chiral vibration for $I\leq 12\hbar$ and
chiral rotation for $I\geq 13\hbar$. In the spin region $15\hbar \leq I \leq 17\hbar$,
one observes that the $P_{R_i}$ distributions are indistinguishable. This provides
the optimal situation for a chiral rotation.


\section{Summary}

In this paper, we have investigated the behavior of the collective rotor in
chiral motion (vibration or rotation) in the particle rotor model by transforming
the rotational wave functions from the $K$-representation to the $R$-representation.
After examining the energy spectra of the doublet bands as well as their
energy differences as functions of the triaxial deformation parameter $\gamma$,
the angular momentum components of the rotor, proton, neutron, and the total system
have been studied in detail at $\gamma=0^\circ$, $10^\circ$, $20^\circ$, and
$30^\circ$. For this purpose, the probability distributions of the rotor
angular momentum ($R$-plots) and is projections onto the three principal
axes ($K_R$-plots) have been calculated and analyzed.

At $\gamma=0^\circ$ and $10^\circ$, the behavior of the rotor in the yrast and yrare
bands is different, and hence the angular momentum geometry does not support
a chiral rotation. At $\gamma=20^\circ$ and $30^\circ$, the evolution of
the collective motion from chiral vibration at low spins to chiral rotation
at high spins has been verified. In the spin region where chiral vibrations occur,
the $P_{R_i}$ distribution for the yrast band has only one peak at $R_i=0\hbar$, while
for the yrare band it has two peaks at nonzero $R_i$. In the spin region where
chiral rotation occurs, the $P_{R_i}$ distributions for the doublet bands are similar,
having two peaks at nonzero $R_i$. Moreover, when the doublet bands become energetically
degenerate, the behavior of the rotor is nearly the same.

To this end, one should note that the $R$-plots and $K_R$-plots presented
in this work are not directly measurable quantities. Therefore, in future
we will use the $R$-plots and $K_R$-plots to calculate and examine the
electromagnetic transition probabilities ($E2$ or $M1$) as fingerprints of
chiral collective motions in triaxially deformed nuclei.

\section*{Acknowledgements}

One of the authors (Q. B. Chen) thanks E. Streck for help in setting up the
numerical codes. Financial support for this work was provided by Deutsche
Forschungsgemeinschaft (DFG) and National Natural Science
Foundation of China (NSFC) through funds provided to the Sino-German CRC 110
``Symmetries and the Emergence of Structure in QCD" (DFG Grant No.
TRR110 and NSFC Grant No. 11621131001), and the National Key R\&D
Program of China (Contract No. 2018YFA0404400 and No. 2017YFE0116700).
The work of Ulf-G. Mei{\ss}ner was also supported by the Chinese Academy
of Sciences (CAS) through a President's International Fellowship Initiative
(PIFI) (Grant No. 2018DM0034) and by the VolkswagenStiftung (Grant No. 93562).


\end{CJK}

\end{document}